\documentclass[12pt]{iopart}

\usepackage[bb=boondox]{mathalfa}
\usepackage[dvipsnames]{xcolor}
\usepackage{braket}
\expandafter\let\csname equation*\endcsname\relax
\expandafter\let\csname endequation*\endcsname\relax
\usepackage{amsmath,amssymb}
\usepackage{tikz}
\usepackage{hyperref}
\allowdisplaybreaks

\newcommand{\ii}{\mathrm{i}}
\newcommand{\ie}{{\it i.e.},\ }
\newcommand{\eg}{{\it e.g.},\ }
\newcommand{\id}{\mathbb{1}} 

\definecolor{dgreen}{rgb}{0,0.6,0}
\definecolor{dred}{rgb}{0.6,0,0}
\definecolor{lred}{rgb}{1,.8,.8}

\begin{document}

\title{Average mutual information for random fermionic Gaussian quantum states}

\author{Lucas Hackl$^{1,2}$, Mario Kieburg$^{1,3,4}$ and Joel Maldonado$^1$\footnote{Corresponding autor}}

\address{$^1$ School of Mathematics and Statistics, The University of Melbourne, Parkville, VIC 3010, Australia}
\address{$^2$ School of Physics, The University of Melbourne, Parkville, VIC 3010, Australia}
\address{$^3$ Fakult\"at f\"ur Physik, Universität Duisburg–Essen, Duisburg, 47057, Germany}
\address{$^4$ Institut Mittag-Leffler, Djursholm, SE-182 60, Sweden}

\ead{lucas.hackl@unimelb.edu.au, m.kieburg@unimelb.edu.au, jmaldonado@student.unimelb.edu.au}

\begin{abstract}
Studying the typical entanglement entropy of a bipartite system when averaging over different ensembles of pure quantum states has been instrumental in different areas of physics, ranging from many-body quantum chaos to black hole evaporation. We extend such analysis to open quantum systems and mixed states, where we compute the typical mutual information in a bipartite system averaged over the ensemble of mixed Gaussian states with a fixed spectrum. Tools from random matrix theory and determinantal point processes allow us to compute arbitrary $k$-point correlation functions of the singular values of the corresponding complex structure in a subsystem for a given spectrum in the full system. In particular, we evaluate the average von Neumann entropy in a subsystem based on the level density and the average mutual information. Those results are given for finite system size as well as in the thermodynamic limit.
\end{abstract}

\section{Introduction}\label{sec:intro}

Entanglement and the associated quantum correlations, which cannot be described by a local classical theory, are hallmarks of quantum theory. The bipartite von Neumann entanglement entropy provides a genuine measure of the amount of quantum entanglement for a pure bipartite quantum state. While it is relatively straightforward to evaluate a given entanglement measure for specific quantum states, a more complex question is what the average and typical behavior of such a measure is when considering a large ensemble of different quantum states. The simplest setting of this type was studied by Don Page in a seminal paper~\cite{page1993average}, where he studied the average bipartite entanglement entropy for Haar-random pure quantum states in a Hilbert space. Since then, this question has been significantly refined with detailed studies of both the average and variance for a large class of pure state ensembles, ranging from states with a centre~\cite{Bianchi:2019stn}, fixed particle number~\cite{garrison2018does,vidmar2017entanglement,liu2018quantum,lydzba2020eigenstate,hackl-volume-law2022}, fermionic Gaussian states~\cite{liu2018quantum,vidmar2018volume,hackl2019average,lydzba2021entanglement,Huang:2022mhd}, bosonic Gaussian states~\cite{fukuda2019typical,iosue2023page,youm2024average} and even states with some fixed non-Abelian charges~\cite{calabrese2021symmetry,majidy2023non,patil2023average,bianchi2024non}. Especially techniques from random matrix theory allowed detailed and quantitative analysis of these questions which went even beyond the computation of the variance~\cite{2019Entrp..21..539W,2020JPhA...53g5302W,Huang:2021btj} and other quantities than the entanglement entropy~\cite{Li:2021fxb,Huang:2021qzt,Wei:2022bed,Huang:2023jmk}. Furthermore, the ensembles of quantum states were not only generated by the Hilbert-Schmidt metric~\cite{PhysRevLett.72.1148,PhysRevE.52.5653,2003JPhA...3610115Z,2004JPhA...37.8457S,2005PhRvA..71c2313Z,2011JMP....52f2201Z,2016PhRvE..93e2106V,2017PhRvE..96b2106W,2020PhRvA.102a2405K} but also by the Bures metric~\cite{2003JPhA...3610083S,2004JPhA...37.8457S,2005PhRvA..71c2313Z,2010JPhA...43e5302O,2016CMaPh.342..151F,2020JPhA...53w5203W,Wei:2020izq,Wei:2021uhj,2022arXiv220803278W,2021PhRvA.104b2438L,2023PhRvE.107c4206L}.

Exploring the question of typical and average quantum correlations for such state ensembles is an interesting question in its own right, but it has also shown to be relevant in other areas of physics beyond quantum information, namely in the context of the black hole information paradox~\cite{page1993information,aurell2024random}, quantum chaos~\cite{leblond2019entanglement,hackl-volume-law2022,liu2023symmetry} and many-body quantum systems~\cite{haque2022entanglement,kliczkowski2023average}. In particular, a number of recent studies~\cite{vidmar2017entanglement,leblond2019entanglement,lydzba2020eigenstate,lydzba2021entanglement,hackl-volume-law2022} have firmly established that the average entanglement entropy of energy eigenstates matches the one of Haar distributed random states at leading order (and even some subleading order in system size) for quantum-chaotic Hamiltonians. In contrast, the average entanglement entropy of energy eigenstates is significantly below the Haar distributed random prediction for integrable models and even matches the behavior for certain fermionic Gaussian states in certain situations, where the Hamiltonian is free, \ie quadratic in creation/annihilation operators.

Fermionic Gaussian states represent a particular suitable ensembles of states for analytical calculation. The reason is that their properties are fully encoded in a simple matrix, the so-called \emph{covariance matrix} or  \emph{complex structure}, whose dimension only grows linearly with the system size (rather than exponentially as the Hilbert space). In particular, the von Neumann entropy of a Gaussian state reduced to a subsystem can be directly computed from the eigenvalues of the restricted covariance matrix to the respective subsystem. Moreover, when defining an ensemble of quantum states we can relate the action of Gaussian transformation directly with a simple group action on the covariance matrix, namely by conjugation with orthogonal matrices. Finally, the statistical properties of the respective eigenvalues can also be studied analytically as the resulting set of covariance matrices forms a known random matrix ensemble~\cite{bianchi2021page,hackl-volume-law2022} described by a so-called \emph{determinental point process}~\cite{log-gases-peter}. The corresponding ensemble is given by random projections of antisymmetric real matrices and pose a special case of the investigations in~\cite{kieburg2019multiplicative} with harmonic analysis techniques introduced and developed in random matrix theory in~\cite{2016arXiv160102586K,2019AIHPB..55...98K,2017arXiv171008794F,2017arXiv171009481K}.

Determinantal point processes are well-studied in random matrix theory as they enable to express all kinds of spectral properties, such as the level density and the $k$-point correlation functions, in terms of a single kernel function~\cite{log-gases-peter}. In the present manuscript, we use these structures to extend the previous studies of pure fermionic Gaussian quantum states towards mixed ones, \ie we consider a mixed Gaussian quantum state with given spectrum for the complex structure and a bi-partition to then apply arbitrary Haar distributed random Gaussian unitary transformations to define an ensemble of quantum states. The average mutual information will be computed for those random quantum states. While we thus still focus on bipartition, though now for mixed states, this can also be readily extended to $n$-partite mutual information measures~\cite{Kumar_2017}.

This manuscript is structured as follows. In Sec.~\ref{section:2}, we review the theory of fermionic Gaussian quantum states, pure as well as mixed, and how their von Neumann entropy and mutual information is expressed in their corresponding complex structure. The latter will be randomised, in Sec.~\ref{sec:rmt-review}, by randomly projecting to the complex structure of a subsystem. The resulting ensembles is a random matrix that exhibits a determinantal point process for which we derive the kernel function. In Sec.~\ref{sec:main-results}, we derive analytical expressions for the von Neumann entropy of the density matrices at finite system size. To illustrate these results we give two examples, namely the bipartition of a mixed two qubit system and the case of maximally degenerate complex structure of the whole system. The main results of the thermodynamic limit and their derivation are collated in Sec.~\ref{sec:thermodynamic.limit}. Therein, we study the case when one subsystem remains of finite size as well as the case when both subsystems scale linearly with the size of the whole system. For the latter, we study the case of maximal degeneracy, separately, because it is the easiest to interpret physically. We conclude in Sec.~\ref{sec:summary-outlook}, where we summarize our findings and give some outlook on open questions and further applications of our results.

\section{Mutual information of fermionic Gaussian quantum states}\label{section:2}

We consider a bi-partite quantum system $\mathcal{H}=\mathcal{H}_A\otimes\mathcal{H}_B$ in a mixed quantum state $\rho:\mathcal{H}\to\mathcal{H}$. We recall that a quantum state is non-negative $\rho>0$, Hermitian $\rho=\rho^\dagger$ and has unit trace $\Tr{\rho}=1$. A state is pure if and only if $\Tr{\rho^2}=1$ which is equivalent to the existence of a vector $\ket{\psi}\in \mathcal{H}$ such that $\rho=\ket{\psi}\bra{\psi}$. 

The bi-partition implies the existence of two reduced density operators
\begin{align}
  \rho^{(A)}=\Tr_B\rho:\ \mathcal{H}_A\to\mathcal{H}_A \qquad{\rm and}\qquad  \rho^{(B)}=\Tr_A\rho:\ \mathcal{H}_B\to\mathcal{H}_B,
\end{align}
where $\Tr_A$ and $\Tr_B$ are the partial traces over $\mathcal{H}_A$ and $\mathcal{H}_B$, respectively. The main question, which is usually addressed, is then whether the state $\rho$ is separable or not. We recall that a quantum state $\rho$ is separable with respect to the bi-partition $\mathcal{H}=\mathcal{H}_A\otimes\mathcal{H}_B$ if one finds probability weights $\{p_j\}_{j\in\mathbb{N}}$ and two sets of quantum states $\{\rho_{j}^{(A)}\}_{j\in\mathbb{N}}$ and $\{\rho_{j}^{(B)}\}_{j\in\mathbb{N}}$ of $\mathcal{H}_A$ and $\mathcal{H}_B$, respectively, so that
\begin{align}
    \rho=\sum_{j=1}^\infty p_j\,\rho_{j}^{(A)}\otimes\rho_{j}^{(B)}.
\end{align}
The question of the existence and finding such probability weights and quantum states is NP-hard. Thus, one resorts to measuring observables that indicate entanglement of $\rho$, the opposite of separability.

In the case of a pure state $\rho$ this  is the entanglement entropy given by
\begin{align}
    S_{\rm ent}=S(\rho^{(A)})=-\Tr(\rho^{(A)}\log\rho^{(A)})=-\Tr(\rho^{(B)}\log\rho^{(B)})=S(\rho^{(B)}).
\end{align}
It is equal to the von Neumann entropy of $\rho^{(A)}$ or equivalently $\rho^{(B)}$. Indeed, for a pure state $\rho$ the spectra of the two reduced  density operators $\rho^{(A)}$ and $\rho^{(B)}$ only differ by their multiplicity of zero eigenvalues. In the following we use the short-hand notation $S_A=S(\rho^{(A)})$ and $S_B=S(\rho^{(B)})$ when directly referring to system $A$ and $B$, respectively.

The situation changes drastically when going over to mixed quantum states. Then, it is usually $S_A\neq S_B$. As the entanglement is a property that should not depend on whether one considers one of the two subsystems it is reasonable to consider quantities such as the mutual information between the subsystems $A$ and $B$,
\begin{align}
    I=S_A+S_B-S(\rho),\label{eq:mutual-information}
\end{align}
which is symmetric in $A$ and $B$. We would like to point out that this observable mixes both, classical and quantum correlations, between the two subsystems. There are more elaborate observables that purely concentrate on quantum correlations, such as entanglement of formation, distillable entanglement or (logarithmic) negativity, see~\cite{bennett1996mixed,wolf2004gaussian,vidal2002computable,plenio2005logarithmic}, though they are harder to study analytically. Thence, we focus on the mutual information alone.

A particular convenient and versatile subclass of quantum states are Gaussian quantum states. We will only  study those for fermionic systems because then the set of all such states is compact, particularly there is a natural uniform measure on the set of all fermionic Gaussian quantum states, see~\cite{hackl-volume-law2022}. Such a natural measure cannot exist for bosonic Gaussian quantum states as they constitute a non-compact manifold, which is even true for a finite number of bosonic modes. To circumvent this problem one usually either resorts to fixing the energy shell~\cite{2006JMP....47j3304L,2011JPhA...44G5207H} or their marginals as it has been recently done in~\cite{2024PhRvL.133f0202A}.

In the fermionic setting, the bipartite splitting of the Hilbert space $\mathcal{H}=\mathcal{H}_A\otimes\mathcal{H}_B$ is given by the splitting of the $N=N_A+N_B$ fermionic modes into the respective subsystems $A$ and $B$. Any fermionic system can be described by fermionic creation and annihilation operators $a_j^{\dagger}$ and $a_j$ with $j=1,\ldots,N$. They satisfy the anticommutation relations
\begin{align}
    \{a_j,a^\dagger_l\}=\delta_{jl} \qquad{\rm and}\qquad \{a_j,a_l\}=\{a^\dagger_j,a^\dagger_l\}=0
\end{align}
for any $j,l=1,\ldots,N$, where $\{X,Y\}=XY+YX$ is the anti-commutator and $\delta_{jl}$ is the Kronecker symbol. For convenience and highlighting the generality of our study, we omit physical dimensions and constants such as the Planck constant.

An equivalent basis of the creation and annilation operators is the Majorana basis consisting of $2N$ Hermitian operators,
\begin{align}
    \gamma_l=\frac{1}{\sqrt{2}}\left( a^\dagger_l + a_l \right), \quad \quad \quad  \gamma_{N+l}=\frac{\ii}{\sqrt{2}}\left ( a^\dagger_l-a_l\right).
\end{align}
They constitute a Clifford algebra with $2N$ generators, meaning they satisfy the anticommutation relations
\begin{align}
    \{\gamma_j,\gamma_l\}=\delta_{jl} \qquad{\rm for\ all}\ j,l=1,\ldots,2N.
\end{align}
We employ the standard representation of the Clifford algebra as Hermitian operators acting on the space $\mathcal{H}=\mathbb{C}^{2^N}$, and analogously $\mathcal{H}_A=\mathbb{C}^{2^{N_A}}$ and $\mathcal{H}_B=\mathbb{C}^{2^{N_B}}$. 

The Majorana basis has several advantages. The most important one for our purposes is that fermionic Gaussian quantum states have a very compact and elegant representation. Indeed, the density operator of such a state can be written as follows
\begin{align}
    \rho_Q=\frac{\exp{\left(- \sum^{2N}_{j,l=1} q_{jl}\gamma_j\gamma_l\right)}}{\Tr\exp{\left(-\sum^{2N}_{j,l=1}q_{jl}\gamma_j\gamma_l\right)}}=\frac{\exp{\left(-\gamma^\dagger Q\gamma\right)}}{\Tr{\exp{\left(-\gamma^\dagger Q\gamma\right)}}},\label{eq:rho_Q}
\end{align}
where the Bogoliubov Hamiltonian $Q=\{ q_{jl}\}_{j,l=1,\ldots,2N}=-Q^T=-Q^*$ is an imaginary antisymmetric $2N\times 2N$ matrix and $\gamma=\left (\gamma_1,\cdots,\gamma_{2N} \right)^\dagger$ is a column vector with operator valued entries $\gamma_l$. The matrix $Q$ corresponds bijectively to a fermionic Gaussian quantum state. This means when fixing $Q$ we have fixed the quantum state $\rho$ and vice versa. This has two important consequences.

Firstly, it helps to study a Hilbert space which is exponentially large in the number of modes $N$ with the help of a matrix space whose dimension only grows algebraically. This feature renders Gaussian quantum states as numerically feasible.

Secondly, the antisymmetric matrix $Q$ opens a way to explicitly parametrise the manifold of all fermionic Gaussian quantum states. For instance, we can quasi-diagonalized any imaginary antisymmetric matrix by an orthogonal transformations $M \in \mathrm{O}(2N)$, 
\begin{align}\label{parametrisation}
     Q=M\Lambda M^{-1}\qquad{\rm with}\qquad \Lambda=\mathrm{diag}(\lambda_1\tau_2,\dots,\lambda_N\tau_2),
\end{align}
where $\lambda_1,\ldots,\lambda_N\geq0$ are the singular values of $Q$ and $\tau_2$ is the second Pauli matrix. When ordering the singular values and restricting $M$ to the coset $\mathrm{O}(2N)/[{\rm SO}^N(2)]$ the parametrisation~\eqref{parametrisation} becomes one-to-one for all matrices $Q$ with non-degenerate  singular values.

In spite of the representation of fermionic Gaussian quantum states in terms of the matrix $Q$ is compact, this matrix is not easily observable. In contrast, the real anti-symmetric $2N\times 2N$ matrix~\cite{hackl-volume-law2022}
\begin{align}
    J=\{J_{jl}\}_{j,l=1,\ldots,2N}\quad{\rm with}\quad J_{jl}=-\ii\,\mathrm{Tr}\big[\rho_Q(\gamma_j\gamma_l-\gamma_l\gamma_j)\big]=2\mathrm{Tr}\big(\rho_Q \gamma_j\gamma_l\big)-\delta_{lj},
\end{align}
which is a symplectic form and known as the complex form as well as the quantum mechanical correlation matrix, can be measured. The bijective relation between the matrices $Q$ and $J$ is simply~\cite{hackl-volume-law2022}
\begin{align}
    J=i\tanh(Q),
\end{align}
which implies for the von Neumann entropy of the corresponding quantum state $\rho_Q$
\begin{align}
    \begin{split}
          S(\rho_Q)&=-\Tr[\rho_Q\log{\rho_Q}]=\frac{1}{2}\Tr s(\ii\,J)\label{eq:entropy_mixed_fermionic}
    \end{split}
\end{align}
where 
\begin{align}
    s(\chi)=- \frac{1+\chi}{2}\log\left(\frac{1+\chi}{2}\right)-\frac{1-\chi}{2}\log\left(\frac{1-\chi}{2}\right).\label{eq:entropy-formula}
\end{align}
When going over to the singular values $y={\rm diag}(y_1,\ldots,y_N)\in[0,1]^N$ of $J$ it becomes explicitly
\begin{align}
    \begin{split}
          S(\rho_Q)&=\sum_{j=1}^N s(y_j),\label{eq:entropy_mixed_fermionic.sum}
    \end{split}
\end{align}
where $y_j=\tanh(\lambda_j)$.

From~\eqref{eq:entropy_mixed_fermionic.sum} it becomes clear that a fermionic Gaussian quantum state is pure if and only if the singular values $y_j=1$ for all $j=1,\ldots,N$. Only then the entropy is vanishing. We underline that this does not mean that the symplectic form $J$ is trivial due to the matrix $M\in{\rm O}(2N)$ it is $J=\ii\,M(\mathbb{1}\otimes\tau_2)M^{-1}$. The latter actually shows that the set of all pure fermionic Gaussian quantum states is isomorphic to the coset ${\rm O}(2N)/{\rm U}(N)$.

Coming back to the bipartite splitting of a mixed quantum state, the mutual information~\eref{eq:mutual-information} depends on the von Neumann entropy  of $\rho_Q$ but also of the reduced density matrices $\rho_Q^{(A)}$ and $\rho_Q^{(B)}$. We note that a quantum state of a subsystem of a fermionic Gaussian quantum state is Gaussian, too. Without loss of generality\footnote{Note that there are some subtleties related to the tensor product for fermionic systems, as discussed in~\cite{szalay2021fermionic}, but for our purpose none of these cause any obstacles.}, we can choose $\gamma_A=(\gamma_1,\ldots,\gamma_{2N_A})^\dagger$ as the Majorana basis in subsystem $A$ and $\gamma_B=(\gamma_{2N_A+1},\ldots,\gamma_{2N})^\dagger$ the one in subsystem $B$. Then, it is
\begin{align}
    J_A=2\Tr_{\mathcal{H}}(\rho_Q\gamma_A\gamma_A^\dagger)-\mathbb{1}_{2N_A}=[J]_A \quad{\rm and}\quad J_B=2\Tr_{\mathcal{H}}(\rho_Q\gamma_B\gamma_B^\dagger)-\mathbb{1}_{2N_B}=[J]_B,
\end{align}
where $\Tr_{\mathcal{H}}$ shall highlight that we only trace over the operator part that acts on the Hilbert space $\mathcal{H}$ and not the indices of $\gamma_j$. Furthermore, $[J]_A$ and $[J]_B$ is a short-hand notation for the $2N_A\times 2N_A$ upper left and $2N_B\times 2N_B$ lower right block of the matrix $J$, respectively. Hence, the mutual information~\eref{eq:mutual-information} is
\begin{align}
    I_{AB}=\frac{1}{2}\Tr s(\ii\,[J]_A)+\frac{1}{2}\Tr s(\ii\,[J]_B)-\frac{1}{2}\Tr s(\ii\,J).
\end{align}

To get a more explicit expression of this observable and setting the stage of our analytical computations, we have to go over to the singular values of $J$, $[J]_A$ and $[J]_B$. The singular values of $J$ are given by $y={\rm diag}(y_1,\ldots,y_N)\in[0,1]^N$. Once these singular values are set, the set of possible singular values of $[J]_A$ and $[J]_B$ are strongly restricted; they are not fixed, though. Their singular values of
\begin{align}
    [J]_A=[M{\rm diag}(\ii\,y_1\tau_2,\ldots,\ii\,y_{N}\tau_2)M^{-1}]_A\quad {\rm and}\quad [J]_B=[M{\rm diag}(\ii\,y_1\tau_2,\ldots,\ii\,y_{N}\tau_2)M^{-1}]_B
\end{align}
also depend on the orthogonal matrix $M\in{\rm O}(2N)$. When choosing $M$ to be randomly Haar distributed on the orthogonal group $M\in{\rm O}(2N)$ we obtain two real antisymmetric random matrices $[J]_A$ and $[J]_B$ despite that the singular values of $J$ are fixed. this carries over to the singular values $x^{(A)}={\rm diag}(x_1^{(A)},\ldots,x_{N_A}^{(A)})$ and $x^{(B)}={\rm diag}(x_1^{(B)},\ldots,x_{N_B}^{(B)})$ as well as to the mutual information
\begin{align}\label{mutual.explicit}
    I_{AB}=\sum_{j=1}^{N_A} s(x_j^{(A)})+\sum_{j=1}^{N_B} s(x_j^{(B)})-\sum_{j=1}^N s(y_j).
\end{align}
While the last term is deterministic for given $y_j$, the distribution of the first two terms is the analytical challenge. To keep the computation at a minimum we are only interested in the mean value of $I_{AB}$. For this purpose, we need to find the joint probability density function (jpdf) and specifically the level density of the singular values of $[J]_A$ and $[J]_B$. Due to the fact that we draw the orthogonal matrices $M$ uniformly from the orthogonal group $\mathrm{O}(2N)$ via the Haar measure, we can equivalently understand this problem as studying the Haar distributed random orthogonal projections of an antisymmetric matrix $J$ with fixed spectrum. This problem has been partially studied in the literature, as we will review in the ensuing sections.

The jpdf of a diagonal subblock of a pure fermionic Gaussian quantum states, \ie when all the singular values are $y_j=1$ and $J=M[\ii\, \mathbb{1}_N\otimes\tau_2]M^{-1}$, have been derived in~\cite{bianchi2021page,hackl-volume-law2022}.  Another special case, which we will discuss in larger detail in Sec.~\ref{sec:case-study}, is a mixed state with the spectrum of $J$ maximally degenerate meaning all the singular values are given by, \ie $y_k=y_0$ and
\begin{align}
   J=M[\ii\, y_0\mathbb{1}_N\otimes \tau_2] M^{-1} \quad{\rm with}\  y_0 \in [0,1].
\end{align}
Evidently, this case is only a rescaled version of the bipartite splitting of a pure fermionic Gaussian quantum state and hence relates to the results in~\cite{bianchi2021page,hackl-volume-law2022}. Yet, this case also illustrates some important effects when starting from a mixed fermionic Gaussian quantum state $\rho_Q$.

\section{Projections of antisymmetric matrices and their spectral statistics}\label{sec:rmt-review}

We proceed by developing the random matrix model and analysing the corresponding spectral statistics at finite matrix dimension. In subsection~\ref{sec:antisymmetric}, we derive the jpdf of the singular values of a random projection of a real antisymmetric matrix by successively applying corank 2 projections and exploiting results of~\cite{kieburg2019multiplicative}. The resulting jpdf exhibits a determinantal point process~\cite{log-gases-peter} whose kernel is addressed in subsection~\eqref{sec:k-point}. Therein, we also relate the average mutual information with this kernel.

\subsection{Random projections of antisymmetric matrices}\label{sec:antisymmetric}

As we stated in Sec.~\ref{section:2}, random Gaussian fermionic states are characterized by the real antisymmetric $2N\times 2N$ matrix $J$. The problem we need to solve is the following. Given the singular values $y={\rm diag}(y_1,\ldots, y_N)\geq0$, what is the distribution of the singular values $x={\rm diag}(x_1,\ldots,x_{m})$ of a random projection of $J$ to the real antisymmetric $2m\times 2m$ matrix 
\begin{align}
    \tilde{J}=[J]_{m}=[M(\ii\,y\otimes\tau_2)M^{-1}]_{m}
\end{align}
with $M\in\mathrm{O}(2N)$ being Haar distributed? This is an orthogonal projection of corank $2(N-m)$. 

The idea is to break the single projection down to $N-m$ successive corank $2$ projections. Then, we can use the result in~\cite[Proposition~A.2]{kieburg2019multiplicative}, which states that given the singular values $x^{(L)}={\rm diag}(x^{(L)}_1,\ldots,x^{(L)}_L)$ the singular values $x^{(L-1)}={\rm diag}(x^{(L)}_1,\ldots,x^{(L-1)}_{L-1})$ of the corank $2$ random projection $[M_L(\ii\,x^{(L)}\otimes\tau_2)M_L^{-1}]_{L-1}$ with $M_L\in\mathrm{O}(2L)$ being Haar distributed is
\begin{align}
\begin{split}\label{jpdf.corank2}
        p^{(L,1)}(x^{(L-1)}|x^{(L)})={}&\frac{(2L-2)!}{(L-1)!} \frac{\Delta_{L-1}([x^{(L-1)}]^2)}{\Delta_{L}([x^{(L)}]^2)} \\
    &\times\det\left[\begin{matrix} 1 \cdots 1 \\ \left(x^{(L)}_k-x^{(L-1)}_j\right)\Theta\left(x^{(L)}_k-x^{(L-1)}_j\right) \end{matrix}\right]_{ \substack{j=1, \cdots ,L-1\\k=1,\cdots,L } }.
\end{split}
\end{align}
We employ the Heaviside step function $\Theta(z)$ which is only unity when $z>0$ and vanishes otherwise and the Vandermonde determinant
\begin{align}\label{Vand}
    \Delta_L(z)=\prod_{1\leq b<c\leq L}(z_c-z_b)=\det[z_b^{c-1}]_{b,c=1,\ldots,L}.
\end{align}
The first row $1 \cdots 1$ of the determinant in~\eqref{jpdf.corank2} has to be read in such a way that every entry of this row is equal to $1$.

We apply equation~\eqref{jpdf.corank2} in a recursive way for the matrix iteration 
\begin{align}
      J^{(l)}=[M_lJ^{(l-1)}M^{-1}_{l}]_{N-l} \quad{\rm and}\quad J^{(0)}=\ii\,y\otimes\tau_2.
\end{align}
with $M_l\in\mathrm{O}(2[N-l+1])$ Haar distributed for all $l=1,\ldots,N-m$. The associated jpdf of the singular values adheres to a recurrence relation, as well, namely
\begin{align}\label{reccurance}
        p^{(N,l)}(x^{(N-l)}|y)=\int_{\mathbb{R}_+^{N-l+1}}{\rm d}[x^{(N-l+1)}]p^{(N-l+1,1)}(x^{(N-l)}|x^{(N-l+1)})p^{(N,l-1)}(x^{(N-l+1)}|y).
\end{align}
The notation ${\rm d}[x^{(N-l+1)}]$ is shorthand for the product of all differentials. The initial condition is $l=1$ which is~\eqref{jpdf.corank2} for $L=N$ and $x^{(L)}=y$.
It happens that this recurrence is fully solved by
\begin{align}
\begin{split}
            p^{(N,l)}(x^{(N-l)}|y)={}&\frac{\Delta_{N-l}([x^{(N-l)}]^2)}{(N-l)!\Delta_N(y^2)} \prod^{N-l-1}_{j=0}\frac{(2l+2j)!}{(2j)!(2l-1)!}  \\
            &\times \det\left[\begin{matrix} y_c^{2a-2}\\ \left(y_c-x^{(N-l)}_b\right)^{2l-1}\Theta\left(y_c-x^{(N-l)}_b\right) \end{matrix}  \right]_{\substack{a=1,\dots,l\\b=1,\dots,N-l\\c=1,\dots,N}}.
\end{split}\label{eq:joint_level_Density_function_2l_projection} 
\end{align}
This can be readily proven by complete induction. 

Surely, Eq.~\eqref{eq:joint_level_Density_function_2l_projection} reduces for $l=1$ to~\eqref{jpdf.corank2} with  $L=N$. For $l>1$ we plug the ansatz~\eqref{eq:joint_level_Density_function_2l_projection} for $l\to l-1$ and~\eqref{jpdf.corank2} for $L=N-l+1$ into~\eqref{reccurance},
\begin{align}
\begin{split}
            p^{(N,l)}(x^{(N-l)}|y)&=\frac{(2[N-l])!}{(N-l)!(N-l+1)!} \left[\prod^{N-l}_{j=0}\frac{(2l+2j-2)!}{(2j)!(2l-3)!}\right] \frac{\Delta_{N-l}([x^{(N-l)}]^2)}{\Delta_N(y^2)}\\
            &\hspace*{-3cm}\times\int_{\mathbb{R}_+^{N-l+1}}{\rm d}[x^{(N-l+1)}]\det\left[\begin{matrix} 1 \cdots 1 \\ \left(x^{(N-l+1)}_k-x^{(N-l)}_j\right)\Theta\left(x^{(N-l+1)}_k-x^{(N-l)}_j\right) \end{matrix}\right]_{ \substack{j=1, \cdots ,N-l\\k=1,\cdots,N-l+1 } }\\
            &\times\det\left[\begin{matrix} y_c^{2a-2}\\ \left(y_c-x^{(N-l+1)}_b\right)^{2l-3}\Theta\left(y_c-x^{(N-l+1)}_b\right) \end{matrix}  \right]_{\substack{a=1,\dots,l-1\\b=1,\dots,N-l+1\\c=1,\dots,N}},
\end{split}
\end{align}
where we cancelled the Vandermonde determinant $\Delta_{N-l+1}([x^{(N-l+1)}]^2)$. In the next step, we apply the generalized Andr\'eief identity~\cite[Appendix C.1]{2010JPhA...43g5201K} and simplify the prefactors, 
\begin{align}
\begin{split}
            p^{(N,l)}(x^{(N-l)}|y)={}&\frac{(2l-2)[(2l-1)(2l-2)]^{N-l}}{(N-l)!} \left[\prod^{N-l-1}_{j=0}\frac{(2l+2j)!}{(2j)!(2l-1)!}\right] \frac{\Delta_{N-l}([x^{(N-l)}]^2)}{\Delta_N(y^2)}\\
            &\times\det\left[\begin{matrix} y_c^{2a-2}\\ \int_0^{y_c}{\rm d}x'\left(y_c-x'\right)^{2l-3} \\ \int_{x_b^{(N-l)}}^{y_c}{\rm d}x'\left(x'-x_b^{(N-l)}\right) \left(y_c-x'\right)^{2l-3} \end{matrix}  \right]_{\substack{a=1,\dots,l-1\\b=1,\dots,N-l\\c=1,\dots,N}}.
\end{split}
\end{align}
The integration of the one-fold integrals in the matrix entries yield the correct terms and one factor of $1/(2l-2)$ and $N-l$ factors of $1/[(2l-2)(2l-1)]$ which cancel with the same prefactors in front of the determinant. This proves the claim.

Summarising, the jpdfs of the singular values of $J_A$ and $J_B$ are given by~\eqref{eq:joint_level_Density_function_2l_projection} for $l=N-N_A=N_B$ and $l=N-N_B=N_A$ as well as setting $x^{(A)}\equiv x^{(N_A)}$ and $x^{(B)}\equiv x^{(N_B)}$, respectively. Those ensembles have particular benefits which we shall exploit below.

The next question we need to address is how we can obtain the level density from these jpdfs. As the computation is the same for both subsystems we consider a corank $2(N-m)$ projection and set $x=x^{(m)}$ to shorten the notation.

\subsection{The \texorpdfstring{$k$}{k}-point correlation functions}\label{sec:k-point}

An important tool when studying spectral statistics of the singular values $x$ is the $k$-point correlation function. These observables are essentially the marginal densities of the jpdf $p^{(N,N-m)}(x|y)$, in particular we integrate over all but $k$ eigenvalues,  
\begin{align}
    R_k^{(N,N-m)}(x_1,\cdots,x_k|y)=\frac{m!}{(m-k)!}\int_{[0,1]^{m-k}} {\rm d}[\lambda]p^{(N,N-m)}(x_1,\cdots,x_k,\lambda_{1},\ldots,\lambda_{m-k}|y).\label{eq:k-point integral}
\end{align}  
We denoted the integration variables differently to emphasise what is integrated out and what is fixed. The prefactor is a standard choice~\cite{log-gases-peter} and becomes convenient when identifying the statistics with a determinantal point process, see below. This, however, means that their normalisation is given by $\int{\rm d}[x]R_k(x_1,\cdots,x_k)=m!/(m-k)!$. 

We are interested in particular averages which can be traced back to the level density. The latter is related to  the $1$-point correlation function like
\begin{align}
    \varrho^{(N,N-m)}(x_1|y)=\frac{1}{N}R_1^{(N,N-m)}(x_1|y).
\end{align} 
The level density is crucial when it comes to the mean values of linear spectral statistics such as the mutual information~\eqref{mutual.explicit}. Say we are interested in the observable $m^{-1}\sum_{j=1}^mf(x_j)$ with some suitably integrable function $f$. Then, it is
\begin{align}
\begin{split}
    \left\langle\frac{1}{m} \sum_{j=1}^mf(x_j)\right\rangle_{m,y}&=\int_{[0,1]^m}{\rm d}[x]\,\frac{1}{m} \sum_{j=1}^mf(x_j)\,p^{(N,N-m)}(x_1,\cdots,x_k,\lambda_{1},\ldots,\lambda_{m-k}|y)\\
    &=\int_0^1{\rm d}x_1 f(x_1)\varrho^{(N,N-m)}(x_1|y)=\frac{1}{m}\int_0^1{\rm d}x_1 f(x_1)R_1^{(N,N-m)}(x_1|y).
\end{split}
\end{align}
The brackets $\langle.\rangle_{m,y}$ indicates the ensemble average while the subscripts shall highlight over what is averaged and the dependence on $y$. For our problem it is for the mutual information~\eqref{mutual.explicit}
\begin{align}\label{mutual.dens}
\begin{split}
    \langle I_{AB}\rangle&=\overbrace{\left\langle \sum_{j=1}^{N_A}s(x_j^{(A)})\right\rangle_{N_A,y}}^{=\langle S_A\rangle}+\overbrace{\left\langle \sum_{j=1}^{N_B}s(x_j^{(B)})\right\rangle_{N_B,y}}^{=\langle S_B\rangle}-\sum_{j=1}^Ns(y_j)\\
    &=\int_0^1 {\rm d}\lambda s(\lambda)\left[R_1^{(N,N_B)}(\lambda|y)+R_1^{(N,N_A)}(\lambda|y)\right]-\sum_{j=1}^Ns(y_j).
\end{split}
\end{align}

The jpdf~\eqref{eq:joint_level_Density_function_2l_projection} is actually a polynomial ensemble on the real antisymmetric matrices, see~\cite[Definition~2.1]{kieburg2019multiplicative}. Those ensembles have a jpdf of the form
\begin{align}
    p(x)=\frac{1}{Z}\Delta_m(x^2)\det[w_c(x_b)]_{b,c=1,\ldots,m}\geq0
\end{align}
with weight functions $w_c$ and the partition function $Z$ which properly normalises the density, \ie $\int{\rm d}[x]p(x)=1$. To see that~\eqref{eq:joint_level_Density_function_2l_projection} indeed admits this form we multiply the $N\times N$ matrix in the numerator by the inverse Vandermonde matrix $V^{-1}$  of $V=\{y_b^{2c-2}\}_{b,c=1,\ldots,N}$ whose matrix entries are
\begin{align}
    \{V^{-1}\}_{bc}=(-1)^{b+c}\frac{\det[y_{c'}^{2b'-2}]_{b'\neq b,c'\neq c}}{\Delta_N(y^2)}=\oint \frac{dz}{2\pi z^{b}}\prod_{j\neq c}\frac{z-y_j^2}{y_c^2-y_j^2},
\end{align}
where the indices $b'$ and $c'$ of the determinant range over $1,\ldots, N$ excluding $b$ and $c$, respectively. The auxiliary contour runs counterclockwise about the origin once and selects the correct cofactor. With the help of this matrix, it is
\begin{align}
\begin{split}
            &p^{(N,N-m)}(x|y)=\frac{\Delta_{m}(x^2)}{m!} \prod^{m-1}_{j=0}\frac{(2N-2m+2j)!}{(2j)!(2N-2m-1)!}  \\
            &\times \det\left[\begin{matrix} \delta_{ac} \\ \displaystyle\sum_{l=1}^N \left(y_l-x_b\right)^{2N-2m-1}\Theta\left(y_l-x_b\right)\oint \frac{dz}{2\pi z^{c}}\prod_{j\neq l}\frac{z-y_j^2}{y_l^2-y_j^2} \end{matrix}  \right]_{\substack{a=1,\dots,N-m\\b=1,\dots,m\\c=1,\dots,N}}\\
            &=\frac{\Delta_{m}(x^2)}{m!} \prod^{m-1}_{j=0}\frac{(2N-2m+2j)!}{(2j)!(2N-2m-1)!}\\
            &\quad\,\,\times \det\left[ \displaystyle\sum_{l=1}^N \Theta\left(y_l-x_b\right)\oint \frac{dz}{2\pi z^{c}}\frac{\left(y_l-x_b\right)^{2N-2m-1}}{ z^{N-m}}\prod_{j\neq l}\frac{z-y_j^2}{y_l^2-y_j^2}   \right]_{b,c=1,\dots,m},
\end{split}\label{jpdf.alter} 
\end{align}
since the first $N-m$ rows can be expanded along the non-zero main diagonal. From this expression we, firstly, see that the ensemble is indeed a polynomial ensemble, and secondly we can read off the corresponding weight functions which are
\begin{align}\label{weight}
\begin{split}
   w_c(\lambda)&= \sum_{l=1}^N\Theta\left(y_l-\lambda\right) \oint \frac{dz}{2\pi\ii z^{c}}\frac{\left(y_l-\lambda\right)^{2N-2m-1}}{ z^{N-m}}\prod_{j\neq l}\frac{z-y_j^2}{y_l^2-y_j^2}\\
   &=\frac{1}{\Delta_N(y^2)}\det[y_b^{2d-2}\ ,\ \left(y_b-\lambda\right)^{2N-2m-1}\Theta\left(y_b-\lambda\right)\ ,\ y_b^{2d'-2}]_{\substack{b=1,\ldots,N\\d=1,\ldots,N-m+c-1\\d'=N-m+c+1,\ldots,N}}.
\end{split}
\end{align}

Polynomial ensembles exhibit determinantal point processes~\cite{kieburg2019multiplicative}, this means that all $k$-point correlation functions can be written in terms of the $k\times k$ determinants
\begin{align}
     R_k^{(N,N-m)}(x_1,\cdots,x_k|y)=\det[K^{(N,N-m)}(x_b,x_c)]_{b,c=1,\ldots,k}
\end{align}
with $K^{(N,N-m)}(x_b,x_c)$ the kernel function which is for all $k=1,\ldots,m$ the same. Without the prefactor in~\eqref{eq:k-point integral} one would find here a combinatorial factor in front of the determinant. The kernel function is, thence, the main object that determines the whole spectral statistics. For a polynomial ensemble it can be always written in the form~\cite{log-gases-peter,kieburg2019multiplicative}
\begin{align}
    K^{(N,N-m)}(x_b,x_c)=\sum_{j=0}^{m-1}p_j^{(N,N-m)}(x_b^2)q_j^{(N,N-m)}(x_c),
\end{align}
where the linear span of $p_0^{(N,N-m)}(\lambda^2),\ldots,p_{m-1}^{(N,N-m)}(\lambda^2)$ is equal to the vector space of even polynomials of degree $2m-2$ and the linear span of $q_0^{(N,N-m)}(\lambda),\ldots,q_{m-1}^{(N,N-m)}(\lambda)$ must agree with the linear span of the weights $w_1(\lambda),\ldots,w_m(\lambda)$. Furthermore, they must be bi-orthonormal, meaning
\begin{align}\label{bi-orthonormality}
    \int_0^1 d\lambda p_j^{(N,N-m)}(\lambda^2)q_l^{(N,N-m)}(\lambda)=\delta_{jl}\quad{\rm for\ all}\ j,l=0,\ldots,m-1.
\end{align}
We underline that the set of bi-orthonormal functions is not unique as any basis of ${\rm span}(w_1(\lambda),\ldots,w_m(\lambda))$ can be chosen which then uniquely determines the set of polynomials that is bi-orthonormal to this choice.

In our case, we choose $q_{c-1}^{(N,N-m)}(\lambda)=w_c(\lambda)$. It happens that then the monomials are bi-orthogonal to these weights  because of
\begin{align}
\begin{split}
    \int_0^1{\rm d}\lambda\,\lambda^{2j-2}w_l(\lambda)&=\frac{1}{\Delta_N(y^2)}\det\left[y_b^{2d-2}\ ,\ \int_0^{y_b}{\rm d}\lambda\,\lambda^{2j-2}\left(y_b-\lambda\right)^{2N-2m-1}\ ,\ y_b^{2d'-2}\right]\\
    &=\frac{1}{\Delta_N(y^2)}\det\left[y_b^{2d-2}\ ,\ \frac{(2j-2)!(2N-2m-1)!}{(2N-2m+2j-2)!}y_b^{2N-2m+2j-2}\ ,\ y_b^{2d'-2}\right]\\
    &=\frac{(2j-2)!(2N-2m-1)!}{(2N-2m+2j-2)!}\delta_{jl},
\end{split}
\end{align}
where the indices of the determinants run over the following sets $b=1,\ldots,N$, $d=1,\ldots,N-m+l-1$ and $d'=N-m+l+1,\ldots,N$. The determinant vanishes for $j\neq l$ because then the $(N-m-l)$th and the $(N-m-j)$th column become proportional. Thence, the corresponding polynomials are
\begin{align}
    p_j^{(N,N-m)}(\lambda^2)=\frac{(2N-2m+2j)!}{(2j)!(2N-2m-1)!}\lambda^{2j}
\end{align}
and the kernel is explicitly
\begin{align}\label{kernel}
\begin{split}
    K^{(N,N-m)}(x_1,x_2)&=\sum_{j=0}^{m-1}\frac{(2N-2m+2j)!}{(2j)!(2N-2m-1)!}\frac{x_1^{2j}}{\Delta_N(y^2)}\\
    &\hspace*{-1.5cm}\times\det[y_b^{2d-2}\ ,\ \left(y_b-x_2\right)^{2N-2m-1}\Theta\left(y_b-x_2\right)\ ,\ y_b^{2d'-2}]_{\substack{b=1,\ldots,N\\d=1,\ldots,N-m+j\\d'=N-m+j+2,\ldots,N}}.
\end{split}
\end{align}
{\color{red} A Christoffel-Darboux formula for this kernel is unlikely to exist as it does not satisfy any obvious recurrence relation.}

The $1$-point correlation is the diagonal kernel function, in particular it is
\begin{align}\label{1-point}
\begin{split}
    R_1^{(N,N-m)}(x_1|y)&=K^{(N,N-m)}(x_1,x_1)=\sum_{j=0}^{m-1}\frac{(2N-2m+2j)!}{(2j)!(2N-2m-1)!}\frac{x_1^{2j}}{\Delta_N(y^2)}\\
    &\hspace*{-1.5cm}\times\det[y_b^{2d-2}\ ,\ \left(y_b-x_1\right)^{2N-2m-1}\Theta\left(y_b-x_1\right)\ ,\ y_b^{2d'-2}]_{\substack{b=1,\ldots,N\\d=1,\ldots,N-m+j\\d'=N-m+j+2,\ldots,N}}.
\end{split}
\end{align}
The expression will be the starting point for the ensuing analysis.

\section{Mutual Information for finite systems}\label{sec:main-results}

We are now ready for deriving explicit expressions for the expected von Neumann entropy of a reduced density matrix. This will be done in subsection~\ref{sec:representations} where we develop various representation which are suitable for performing the thermodynamics limits. To illustrate the results we discuss the example of a bi-partite two qubit system in subsection~\ref{sec:example.1} and the case of maximal degeneracy of the singular value spectrum of the complex structure $J$ in subsection~\ref{sec:example.2}.

\subsection{Analytical representations of the von Neumann entropy}\label{sec:representations}

In order to compute the average mutual information~\eqref{mutual.dens} we need the average von Neumann entropy of the two subsystems. The average entropy density of a subsystem of size $m$ is
\begin{align}
\begin{split}
\langle S(\rho^{(m)})\rangle_y&=\left\langle\sum_{j=1}^ms(x_j)\right\rangle_{m,y}=\int_0^1{\rm d}x \,s(x)  R_1^{(N,N-m)}(x|y).
\end{split}
\end{align}
When plugging in~\eqref{eq:entropy-formula} and~\eqref{1-point} and interchanging the integral with the sum we arrive at
\begin{align}\label{av.entropy}
\begin{split}
\langle S(\rho^{(m)})\rangle_y&=m\log(2)\\
&\hspace*{-1.cm}-\sum_{j=0}^{m-1}\frac{1}{\Delta_N(y^2)}\det[y_b^{2d-2}\ ,\ y_b^{2(N-m+j)}\widetilde{G}_{N-m,j}(y_b^2)\ ,\ y_b^{2d'-2}]_{\substack{b=1,\ldots,N\\d=1,\ldots,N-m+j\\d'=N-m+j+2,\ldots,N}}\\
&=m\log(2)-\sum_{j=0}^{m-1}\sum_{l=1}^N\oint\frac{{\rm d}z}{2\pi \ii z}\left(\frac{y_l^2}{z}\right)^{N-m+j}\widetilde{G}_{N-m,j}(y_l^2)\prod_{k\neq l}\frac{z-y_k^2}{y_l^2-y_k^2}
\end{split}
\end{align}
with the function
\begin{align}
\begin{split}
\widetilde{G}_{N-m,j}(\lambda^2)&=-\frac{(2N-2m+2j)!}{(2j)!(2N-2m-1)!}\lambda^{-2(N-m+j)}\int_0^{\lambda}{\rm d}x\,s(x)x^{2j}(\lambda-x)^{2N-2m-1}+\log(2)\\
={}&\frac{(2N-2m+2j)!}{(2j)!(2N-2m-1)!}\int_0^1{\rm d} x\\
&\times\left[\frac{1+\lambda x}{2}\log\left(1+\lambda x\right)+\frac{1-\lambda x}{2}\log\left(1-\lambda x\right)\right]x^{2j}(1-x)^{2N-2m-1},
\end{split}
\end{align}
where we rescaled $x\to\lambda x$ in the second step. We split off the term $m\log(2)$ as it is the maximal entropy that a fermionic system with $m$ modes can have while the second term is the deviation from that.

The integral for $\widetilde{G}_{N-m,j}(\lambda^2)$ can be computed by Taylor-expanding the integrand in $\lambda$, \ie
\begin{align}
\begin{split}
\frac{1+\lambda x}{2}\log\left(1+\lambda x\right)+\frac{1-\lambda x}{2}\log\left(1-\lambda x\right)&=-\sum_{l=1}^\infty\frac{1}{l}\left[\frac{1+\lambda x}{2}(-\lambda x)^l+\frac{1-\lambda x}{2}(\lambda x)^l\right]\\
&=-\frac{1}{2}\sum_{l=2}^\infty\left(\frac{1}{l}-\frac{1}{l-1}\right)\left[(-\lambda x)^l+(\lambda x)^l\right]\\
&=\sum_{o=1}^\infty\frac{(\lambda x)^{2o}}{2o(2o-1)}.
\end{split}
\end{align}
This series is uniformly convergent for any $\lambda x\in[0,1]$ so that we can exchange it with the integral. In this way, we can perform the integration which are beta functions leading to
\begin{align}
\begin{split}
\widetilde{G}_{N-m,j}(\lambda^2)&=\sum_{o=1}^\infty\frac{(2N-2m+2j)!(2j+2o)!}{(2N-2m+2j+2o)!(2j)!} \frac{\lambda ^{2o}}{2o(2o-1)}. 
\end{split}
\end{align}
This function is essentially the difference of two hypergeometric functions. It has always a well-defined limit regardless how $N$ and $m$ tend to infinity or whether the summing index $j$ does, too. The reasons for the existence of the limits are $\lambda\in[0,1]$ as well as $j<m$.

We are not finished with massaging the expression due to the determinant~\eqref{av.entropy} we can subtract multiples of the last $m-1-j$ columns from the $(m-j)$th last one so that the first $(m-1-j)$th powers in $\widetilde{G}_{N-m,j}$ are cancelled. Moreover, we reflect $j\to m-1-j$ as it simplifies some expressions. Defining
\begin{align}\label{G.def}
\begin{split}
G_{N,m}(z,z')&=\sum_{j=0}^{m-1}\sum_{o=0}^\infty\frac{(2N-2-2j)!(2m+2o)!}{(2m-2-2j)!(2N+2o)!} \frac{z^j{z'}^{o}}{(2o+2j+2)(2o+2j+1)},
\end{split}
\end{align}
we obtain
\begin{align}\label{av.entropy.b}
\begin{split}
\langle S(\rho^{(m)})\rangle_y&=m\log(2)-\sum_{l=1}^N\oint\frac{{\rm d}z}{2\pi \ii }\frac{y_l^{2N}}{z^{N}}G_{N,m}(z,y_l^2)\prod_{k\neq l}\frac{z-y_k^2}{y_l^2-y_k^2}.
\end{split}
\end{align}
Of course, we can evaluate the contour integral using $\oint z^d dz/(2\pi \ii z)=\delta_{0,d}$ and the generating function of the elementary symmetric polynomials
\begin{equation}
   \prod_{k\neq l}(z-y_k^2)=\sum_{n=1}^{N}(-1)^{n-1}e_{n-1}(y^2_{\neq l})z^{N-n},
\end{equation}
where $y^2_{\neq l}$ is the set of $y_k^2$ with $k\neq l$.
Evaluating the Kronecker delta symbol via $n=j-1$, this leads to
\begin{align}
    \braket{S(\rho^{(m)})}_y&=m \log(2)+\sum^{N}_{l=1}\sum^{m}_{n=1}\sum^{\infty}_{o=0}\tfrac{(-1)^n(2N-2n)!(2m+2o)!y_l^{2N+2o}}{2(o+n)(2o+2n-1)(n-N-o)(2m-2j)!(2N+2o)!}\frac{e_{n-1}(y^2_{\neq l})}{\prod_{k\neq l}(y_l^2-y_k^2)}.\label{eq:explicit-sum}
\end{align}
While expression~\eref{eq:explicit-sum} provides an explicit sum that could be probably used for numerical evaluation, we aim to make analytical progress and therefore instead continue with~\eref{av.entropy.b}.

In the next step, we replace the sum over $l$ by another contour integral over the unit circle ($|z'|=1$)
\begin{align}\label{av.entropy.c}
\begin{split}
\langle S(\rho^{(m)})\rangle_y&=m\log(2)-\oint\frac{{\rm d}z}{2\pi \ii }\oint\frac{{\rm d}z'}{2\pi \ii (z-z')}\left(\frac{z'}{z}\right)^{N}G_{N,m}(z,z')\prod_{k=1}^N\frac{z-y_k^2}{z'-y_k^2}.
\end{split}
\end{align}
We need to ensure that the radius of the circular $z$-contour satisfies $|z|>1$ so that only the residues at $z'=y_k^2$ contribute but not at the pole $z'=z$.

The function $G_{N,m}(z,z')$ has also a more suitable representation than the series which would allow an analytic continuation beyond the unit circle. For the sum over $o$ we rewrite
\begin{align}
\begin{split}
&\sum_{o=0}^\infty\frac{(2m+2o)!(2N-2m-1)!}{(2N+2o)!} \frac{{z'}^{o}}{(2o+2j+2)(2o+2j+1)}\\
&=\int_0^1{\rm d}\chi\int_0^1{\rm d}\eta\frac{(1-\chi)^{2N-2m-1}\chi^{2m}(1-\eta)\eta^{2j}}{1-z'\chi^2\eta^2},
\end{split}
\end{align}
which results from employing the Beta function twice and the geometric series which converges uniformly in $\chi,\eta\in[0,1]$ for any $|z'|<1$. From this result we can see that for $j\geq 1$ the behaviour for $|z'|\to\infty$ is of the order $\mathcal{O}(1/z')$ while for $j=0$ it is  $\mathcal{O}(\log(z')/z')$. Thence, the latter asymptotic behaviour  is shared with $G_{N,m}(z,z')$. This allows us to deform the contour of $z'$ to a circular contour with an infinite radius in~\eqref{av.entropy.c} where it will vanish as the integrand drops off like $\mathcal{O}(\log(z')/(z')^2)$. There are, however, two  additional contributions. The first is the residue at $z'=z$ as we swap the two contours, which also allows us to bring the contour  of $z$ to a circular one  with radius $|z|<1$. The second is a branch cut of $G_{N,m}(z,z')$ along $z'\in[1,\infty)$ which is given by ($t\geq1$)
\begin{align}
\begin{split}
2\pi \ii F_{N,m}(z,t)={}&\lim_{\varepsilon\to0}[G_{N,m}(z,t+\ii \varepsilon)-G_{N,m}(z,t-\ii \varepsilon)]\\
={}&2\pi\ii\sum_{j=0}^{m-1}\frac{(2N-2-2j)!z^j}{(2m-2-2j)!(2N-2m-1)!}\int_0^1{\rm d}\chi\int_0^1{\rm d}\eta\\
&\times(1-\chi)^{2N-2m-1}\chi^{2m-2}(1-\eta)\eta^{2j-2}\delta(\chi^{-2}\eta^{-2}-t)\\
={}&\frac{\pi\ii}{t}\int_{1/\sqrt{t}}^1{\rm d}\chi(1-\chi)^{2N-2m-1}\chi^{2m-2}(\sqrt{t}\chi-1)\\
&\times\sum_{j=0}^{m-1}\frac{(2N-2-2j)!}{(2m-2-2j)!(2N-2m-1)!}\left(\frac{z}{t\chi^2}\right)^{j}.
\end{split}
\end{align}
We applied here the Sokhotski–Plemelj theorem to obtain the Dirac delta function $\delta(\chi^{-2}\eta^{-2}-t)$.

After the above contour deformations, the entropy becomes
\begin{align}\label{av.entropy.d}
\begin{split}
\langle S(\rho^{(m)})\rangle_y&=m\log(2)-\oint\frac{{\rm d}z}{2\pi \ii }G_{N,m}(z,z)\\
&\quad\,\,-\oint\frac{{\rm d}z}{2\pi \ii }\int_1^\infty\frac{{\rm d}t}{z-t}\left(\frac{t}{z}\right)^{N}F_{N,m}(z,t)\prod_{k=1}^N\frac{z-y_k^2}{t-y_k^2}.
\end{split}
\end{align}
The function $G_{N,m}(z,z)$ is holomorphic inside the unit disc so that the contour integral over a circle with a radius $|z|<1$ vanishes, \ie
\begin{align}\label{av.entropy.d1}
\begin{split}
\langle S(\rho^{(m)})\rangle_y&=m\log(2)-\oint\frac{{\rm d}z}{2\pi \ii }\int_1^\infty\frac{{\rm d}t}{z-t}\left(\frac{t}{z}\right)^{N}F_{N,m}(z,t)\prod_{k=1}^N\frac{z-y_k^2}{t-y_k^2}.
\end{split}
\end{align}
 For $F_{N,m}(z,t)$ we, additionally, rewrite the sum over $j$ in terms of a contour integral  as follows
\begin{align}
\begin{split}
&\sum_{j=0}^{m-1}\frac{(2N-2-2j)!}{(2m-2-2j)!(2N-2m-1)!}\left(\frac{z}{t\chi^2}\right)^{j}\\
&=2(N-m)\oint\frac{{\rm d}\eta}{2\pi\ii }\frac{1}{\eta^{2m-1}(1-\eta)^{2N-2m+1}(1-\eta^2z/[t\chi^2])},
\end{split}
\end{align}
where we choose a circular contour with radius $|\eta|<1$ so that the residue is only taken at $\eta=0$.
Summarising everything, we arrive  at an expression for the entropy of the form
\begin{align}\label{av.entropy.e}
\begin{split}
\langle S(\rho^{(m)})\rangle_y&=m\log(2)-(N-m)\oint\frac{{\rm d}z}{2\pi \ii }\int_1^\infty {\rm d}t\int_{1/\sqrt{t}}^1{\rm d}\chi\oint\frac{{\rm d}\eta}{2\pi\ii }\\
&\hspace*{-1cm}\times\frac{\eta(\sqrt{t}\chi-1)}{(z-t)(1-\eta)(1-\chi)(t\chi^2-\eta^2z)}\left(\frac{1-\chi}{1-\eta}\right)^{2N-2m}\left(\frac{\chi}{\eta}\right)^{2m}\prod_{k=1}^N\frac{1-y_k^2/z}{1-y_k^2/t}.
\end{split}
\end{align}

For the last final adjustment, we interchange the $t$ and the $\chi$ integral while respecting $1/\sqrt{t}<\chi$ implying $1/\chi^2<t$ and afterwards rescale $t\to t/\chi^2$,
\begin{align}\label{av.entropy.f}
\begin{split}
\langle S(\rho^{(m)})\rangle_y&=m\log(2)-(N-m)\oint_{|z|=1/2}\frac{{\rm d}z}{2\pi \ii }\int_{0}^1{\rm d}\chi\int_1^\infty{\rm d}t\oint_{|\eta|=1/2}\frac{{\rm d}\eta}{2\pi\ii }\\
&\quad\,\,\times\frac{\eta(\sqrt{t}-1)}{(1-\eta)(1-\chi)(\chi^2 z-t)(t-\eta^2 z)}\left(\frac{1-\chi}{1-\eta}\right)^{2N-2m}\left(\frac{\chi}{\eta}\right)^{2m}\prod_{k=1}^N\frac{1-y_k^2/z}{1-\chi^2y_k^2/t}\\
&=m\log(2)+(N-m)\int_{0}^1{\rm d}\chi\int_1^\infty\frac{{\rm d}t}{t}\oint_{|\eta|=1/2}\frac{{\rm d}\eta}{2\pi\ii }\frac{\eta(\sqrt{t}-1)}{(1-\eta)(1-\chi)(\chi^2-\eta^2) }\\
&\quad\,\,\times\left(\frac{1-\chi}{1-\eta}\right)^{2N-2m}\left(\frac{\chi}{\eta}\right)^{2m}\left[1-\prod_{k=1}^N\frac{1-\eta^2y_k^2/t}{1-\chi^2y_k^2/t}\right].
\end{split}
\end{align}
In the second equality, we have used the fact that the only poles in the exterior of the $z$-contour lie at $z=t/\chi^2$ and $z=t/\eta^2$ so that we can apply the residue theorem.

The two representations~\eqref{av.entropy.b} and~\eqref{av.entropy.f} for the von Neumann entropy will be useful when it comes to asymptotic analysis for large systems. The structure of~\eqref{av.entropy.f} even highlights a supersymmetry behind this quantity which is indeed well-known to exist for the generating functions of the $k$-point correlation functions, \eg see~\cite[Section 6.1]{2011JPhA...44B5210K}. We will not delve deeper into this, here.

\subsection{Example 1: Smallest non-trivial system}\label{sec:example.1}

\begin{figure}[t!]\centering
\includegraphics[width=8cm]{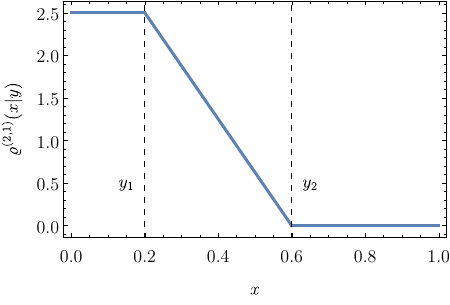}
\caption{Sketch of the level density $\varrho(x)=\varrho^{(2,1)}(x|y)$ for Example 1 following~\eref{Gaussian Fermionic States 2 case piece wise} for a system of two fermionic modes with singular values $(y_1,y_2)$ and a subsystem of a single fermionic mode. For the illustration, we chose $y_1=0.2$ and $y_2=0.6$.
}\label{fig:realizationsof the level density}
\end{figure}

The smallest non-trivial system consists of two qubits and a subsystem of a single fermionic mode. By equation~\eqref{1-point}, we compute the level density for $0\leq y_1\leq y_2\leq 1$ as
\begin{align}
     \varrho^{(2,1)}(x|y)=R_1^{(2,1)}(x|y)&=\frac{2}{(y_2^2-y_1^2)}\det\left[\begin{matrix}
            1 && (y_1-x)\Theta(y_1-x)\\
            1 && (y_2-x)\Theta(y_2-x)
        \end{matrix}\right]\nonumber\\
     &=\begin{cases} 
      \displaystyle 2\frac{(y_2-y_1)}{(y^2_2-y^2_1)}, & x\leq y_1,\\[2mm]
      \displaystyle 2\frac{(y_2-x)}{(y^2_2-y^2_1)}, & y_1 \leq x \leq y_2,\\[2mm]
      0, & \text{else}.
   \end{cases}\label{Gaussian Fermionic States 2 case piece wise}
\end{align}
We illustrate this function in Fig.~\ref{fig:realizationsof the level density}. In the limit $y_1\to y_2$, the level density approaches a step function.

\begin{figure}[t]
    \begin{tikzpicture}
    \draw (0,0) node[inner sep=0pt]{\includegraphics[width=\linewidth]{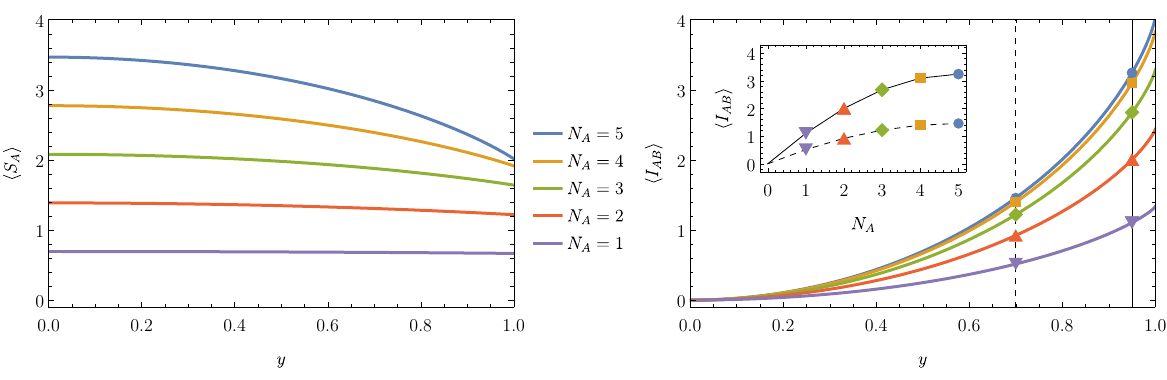}};
    \draw (-5,1.9) node[scale=.7]{\textbf{(a)}} (2,1.9) node[scale=.7]{\textbf{(b)}};
    \end{tikzpicture}
    \caption{
    \textbf{(a)} The average von Neumann entropy $\braket{S_A}=\braket{S(\rho_A)}$ in a subsystem of a single fermionic mode is shown that is embedded in a system with two fermionic modes and a Gaussian state whose covariance matrix has singular values $y_1,y_2\in[0,1]$. The case $y_1=y_2=1$ corresponds to a pure Gaussian state, while $y_1=y_2=0$ corresponds to the maximally mixed state. \textbf{(b)} This is the mutual information $I_{AB}$ corresponding to the setup in (a).
    \label{fig:illustration-1-in-2-modes}}
\end{figure}

We can consider the average entanglement entropy by evaluating
\begin{align}
    \langle S_A\rangle=\langle S_B\rangle=\int_0^1 {\rm d}x \varrho^{(2,1)}(x|y) s(x)=\frac{\widehat{G}(y_1)-\widehat{G}(y_2)}{y_2^2-y^2_1}\,,\label{eq:average-entropy}
\end{align}
where the relevant integrals could be executed piecewise and the function $\widehat{G}(y)$ is
\begin{align}
\begin{split}
    \widehat{G}(y)&=2\int_0^y {\rm d}x\,s(x)(y-x)=2\sum_{o=0}^\infty\frac{(2o)!}{(2o+4)!} y ^{2o+4}-\log(2)y^2\\
    &=\frac{1}{6}\left[(1+3y^2)\log(1-y^2)+(3y+y^3)\log\left(\frac{1+y}{1-y}\right)-(5+\log{64})y^2\right].
\end{split}
\end{align}
Even though, the expression~\eref{Gaussian Fermionic States 2 case piece wise} is not invariant under $y_1\leftrightarrow y_2$, the resulting average entropy~\eqref{eq:average-entropy} is manifestly symmetric, so that we do not need to replace $y_1\to \min(y_1,y_2)$ and $y_2\to \max(y_1,y_2)$ when plotting it for arbitrary $0\leq y_i\leq 1$. Due to this symmetry of the system, we have $\braket{S_A}=\braket{S_B}$, so the mutual information is given by
\begin{align}
    \braket{I_{AB}}=2\braket{S_A}-s(y_1)-s(y_2)\,.
\end{align}
The average von Neumann entropy $\braket{S_A}$ and mutual information $\braket{I_{AB}}$ as a function of $y_1$ and $y_2$ are illustrated in Fig.~\ref{fig:illustration-1-in-2-modes}.

\subsection{Example 2: Maximal degeneracy of \texorpdfstring{$J$}{J}'s spectrum}\label{sec:example.2}

In the situation of maximal degeneracy $y_k=y\in[0,1]$ for all $k=1,\ldots,N$ for a finite number of fermions $N$ the von Neumann entropies $S_A$ and $S_B$ are related. The reason is that the spectra of the two sub-blocks $[J]_A$ and $[J]_B$ of the complex structure $J=y\hat{\tau}_2$ are related. To see this we introduce the orthogonal projections $\Pi_A$ and $\Pi_B$ such that $[J]_A=\Pi_A^T J\Pi_A$ and $[J]_B=\Pi_B^T J\Pi_B$ and consider the characteristic polynomial
\begin{equation}
    D(x)=\det[x\id_{2N_A}-\Pi_A^T J\Pi_A]=x^{2N_A}\det\left[\id_{2N}-\frac{y}{x}\Pi_A\Pi_A^T \hat{\tau}_2\right].
\end{equation}
As we deal with orthogonal projections, it is $\Pi_A\Pi_A^T=\id_{2N}-\Pi_B\Pi_B^T$ leading to
\begin{equation}
\begin{split}
    D(x)&=x^{2N_A}\det\left[\id_{2N}-\frac{y}{x}\hat{\tau}_2+\frac{y}{x}\Pi_B\Pi_B^T \hat{\tau}_2\right]\\
    &=\frac{(x^2-y^2)^N}{x^{2N_B}}\det\left[\id_{2N}+\Pi_B\Pi_B^T \left(\frac{y^2}{x^2-y^2}\id_{2N}+\frac{xy}{x^2-y^2}\hat{\tau}_2\right)\right],
\end{split}
\end{equation}
where we employed
\begin{equation}
    \left[\id_{2N}-\frac{y}{x}\hat{\tau}_2\right]^{-1}=\frac{x^2}{x^2-y^2}\id_{2N}+\frac{xy}{x^2-y^2}\hat{\tau}_2.
\end{equation}
Next we use $\Pi_B^T\Pi_B=\id_{2N_B}$ as well as $\hat\tau_2=-\hat\tau_2^T$ to arrive at
\begin{equation}
\begin{split}
    D(x)&=\frac{(x^2-y^2)^N}{x^{2N_B}}\det\left[\frac{x^2}{x^2-y^2}\id_{2N_B}+\frac{xy}{x^2-y^2}\Pi_B^T\hat{\tau}_2\Pi_B\right]\\
    &=(x^2-y^2)^{N_A-N_B}\det\left[x\id_{2N_B}-y\Pi_B^T\hat{\tau}_2\Pi_B\right].
\end{split}
\end{equation}
Therefore, when $x$ is a singular value of $[J]_A$ then it is also one of $[J]_B$ with the very same multiplicity unless it is $x=y$ where the difference of multiplicity between $[J]_A$ and $[J]_B$ is $|N_A-N_B|$. This leads to the following relation between the two von Neumann entropies
\begin{equation}\label{ent.rel.max}
    S_A=S_B+(N_A-N_B)s(y)
\end{equation}
which means for the mutual information
\begin{equation}\label{mutual.max}
    I_{AB}=S_A+S_B-Ns(y)=2S_A-2N_As(y).
\end{equation}

When computing the expectation value we may use the result~\eqref{av.entropy.f}. In the present case, that one simplifies to
\begin{align}\label{finite.max.degeneracy.a}
\begin{split}
\langle S(\rho^{(m)})\rangle_y={}&m\log(2)+(N-m)\int_{0}^1{\rm d}\chi\int_1^\infty\frac{{\rm d}t}{t}\oint_{|\eta|=1/2}\frac{{\rm d}\eta}{2\pi\ii }\frac{\eta(\sqrt{t}-1)}{(1-\eta)(1-\chi)(\chi^2-\eta^2) }\\
&\times\left(\frac{1-\chi}{1-\eta}\right)^{2N-2m}\left(\frac{\chi}{\eta}\right)^{2m}\left[1-\left(\frac{1-\eta^2y^2/t}{1-\chi^2y^2/t}\right)^N\right].
\end{split}
\end{align}
An alternative and faster way would be to realise that the level density of singular values of $[J_A]$ is the same for a subsystem of a uniformly distributed pure Gaussian fermionic quantum state with the same system and subsystem size only rescaled by $y$. Especially the following holds true for the resulting $k$-point correlation function,
\begin{align}
    R_k^{(N,N_A)}(x_1,\dots,x_k|y\id_{N})=\frac{1}{y^k}R_k^{(N,N_A)}\left(\left.\frac{x_1}{y},\dots,\frac{x_k}{y}\right|\id_{N}\right).
\end{align}
In~\cite{bianchi2021page} two of the authors have discussed the situation of a pure quantum state, which was also based on the result~\cite[Proposition A.2]{kieburg2019multiplicative}. In this case one could trace the problem back to a particular Jacobi ensemble. For $N_A\leq N_B$, which can be always chosen in the case of a pure quantum state without loss of generality, the corresponding kernel $\widetilde{K}(x_1,x_2)$ is
\begin{align}\label{Jacobi-kernel}
    \widetilde{K}(x_1,x_2)=\sum^{N_A-1}_{j=0}\psi_j(x_1)\psi_j(x_2)\quad\text{with}\quad \psi_j(x)=\frac{(1-x^2)^{\frac{N_B-N_A}{2}}}{\sqrt{c_j}}\mathcal{P}^{(N_B-N_A,N_B-N_A)}_{2j}(x),
\end{align}
where $\mathcal{P}^{(\alpha,\beta)}_k(x)$ refers to the Jacobi polynomial, and the normalization constant $c_j$ is
\begin{align}
    c_j=\frac{2^{2N_B-2N_A}[(2j+N_B-N_A)!]^2}{(2j)!(2j+2N_B-2N_A)!(4j+2N_B-2N_A+1)}.
\end{align}

The kernel $\widetilde{K}(x_1/y,x_2/y)/y$ is certainly related to the kernel we have found in~\eqref{kernel}. To recognise this, we highlight that all $k$-point correlation functions are invariant by multiplying the kernel with co-cycles, \ie $K(x_1,x_2)\to K(x_1,x_2)g(x_1)/g(x_2)$. In the present case, to go over from~\eqref{kernel} to~\eqref{Jacobi-kernel} one needs to choose $g(x_1)=(1-x_1^2/y^2)^{(N_B-N_A)/2}$. Moreover, the basis of bi-orthonormal functions is not uniquely given. Choosing a different basis for the set of polynomials requires another basis for the weights. In the general case of a mixed quantum state, the basis for the polynomials had been the monomials instead of the Jacobi polynomials. The most important property is the satisfaction of the bi-orthonormality relation~\eqref{bi-orthonormality} and that they linearly span the same vector spaces.  The latter is obvious for the polynomials but also for the weights~\eqref{weight} this can be readily checked.


\begin{figure}[t]
    \begin{tikzpicture}
    \draw (0,0) node[inner sep=0pt]{\includegraphics[width=\linewidth]{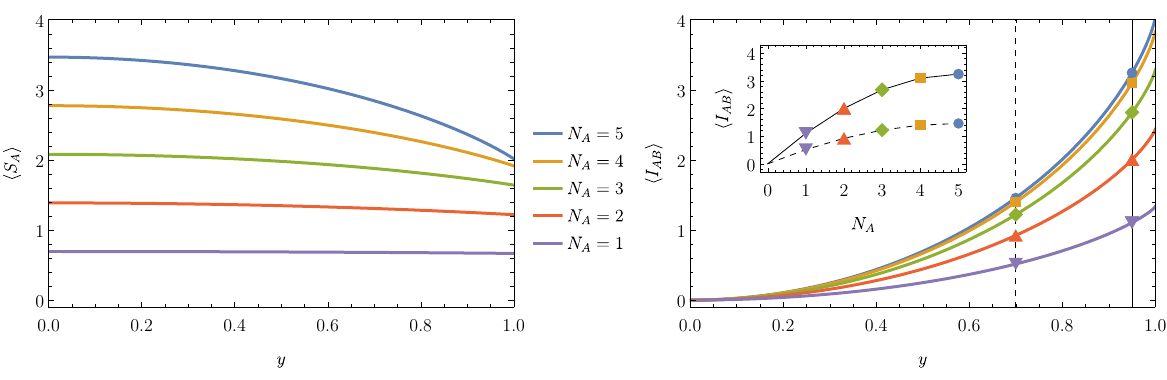}};
    \draw (-7.35,1.8) node[scale=.7]{\textbf{(a)}} (1.15,1.8) node[scale=.7]{\textbf{(b)}};
    \end{tikzpicture}
    \caption{
    \textbf{(a)} Average von Neumann entropy $\braket{S_A}=\braket{S(\rho_A)}$ in a subsystem of $N_A$ fermionic modes embedded in a system with $N=10$ fermionic modes for a Gaussian quantum state whose covariance matrix is maximally degenerate with a singular value $y\in[0,1]$. \textbf{(b)} The corresponding mutual information for the same setting as in (a). The inset shows the mutual information as a function of the subsystem size, \ie the ``Page curve'' of the mutual information for $y=0.7$ (dashed) and $y=0.95$ (solid).
    \label{fig:SA-and-IAB-max-degenerate}}
\end{figure}

The average von Neumann entropy in a subsystem and the mutual information of a subsystem and its complement is also immediate. For $N_A\leq N_B$ it is
\begin{equation}\label{max.deg.rel}
    \braket{S_A}=\int_0^y{\rm d}x\,\varrho^{(N,N_B)}(x)s(x)=\int_0^1{\rm d}x\,\widetilde{K}(x,x)s(yx)
\end{equation}
and using~\eref{eq:mutual-information}. The explicitly resulting expressions are not particularly insightful so that we decided to only show their resulting functional forms in Fig.~\ref{fig:SA-and-IAB-max-degenerate}. The  von Neumann entropy decreases for larger $y$ as expected, while the mutual information increases. The reason is that for small $y$ the state approaches the maximally mixed state, which is also maximally mixed in each subsystem and even closer to that. Thus, the limit $y\to0$ maximizes the von Neumann entropy but at the same time does not encode any correlations between the two subsystems so that the mutual information vanishes.

\section{Statistics in the limit of large systems}\label{sec:thermodynamic.limit}

There are several thermodynamical limits ($N\to\infty$) of the von Neumann entropies and mutual information one can study. The most important ones are those where either one of the subsystems has a finite system size, see subsection~\ref{sec:finite.subsystem}, or where both subsystems scale linearly with the total system size, subsection~\ref{sec:large.subsystem}. Anew, we will study the case of maximal spectral degeneracy of the complex structure $J$, separately, in subsection~\ref{sec:case-study}, since its results are the simplest to interpret.

\subsection{Finite subsystem \texorpdfstring{$A$}{A}}\label{sec:finite.subsystem}

When taking the limit $N\to\infty$ but keeping $N_A$ fixed, the evaluation of the entropy of the reduced density matrix $\rho^{(A)}$ becomes tremendously simpler. We start with the integral representation in~\eqref{av.entropy.b} for $m=N_A$,
\begin{align}
\begin{split}
\langle S_A\rangle&=N_A\log(2)-\sum_{l=1}^N\oint\frac{{\rm d}z}{2\pi \ii }\frac{y_l^{2N}}{z^{N}}G_{N,N_A}(y_l^2)\prod_{k\neq l}\frac{z-y_k^2}{y_l^2-y_k^2}.
\end{split}
\end{align}
and notice the simplification
\begin{align}
\begin{split}
G_{N,N_A}(y_l^2)&\sim\sum_{j=0}^{N_A-1}\sum_{o=0}^\infty\frac{(2N_A+2o)!}{(2N_A-2-2j)!} \frac{(2N)^{-2-2j-2o}z^jy_l ^{2o}}{(2o+2j+2)(2o+2j+1)}.
\end{split}
\end{align}
Since $y_l\in[0,1]$, the leading order term comes from $j=o=0$ so that
\begin{align}
\begin{split}
\langle S_A\rangle&\sim N_A\log(2)-\frac{(2N_A)!}{8N^2(2N_A-2)!}\sum_{l=1}^N\oint\frac{{\rm d}z}{2\pi \ii }\frac{y_l^{2N}}{z^{N}} \prod_{k\neq l}\frac{z-y_k^2}{y_l^2-y_k^2}\\
&\sim N_A\log(2)-\frac{N_A(2N_A-1)}{4N^2}\sum_{l=1}^N\frac{y_l^{2N}}{\prod_{k\neq l}(y_l^2-y_k^2)}.
\end{split}
\end{align}
The sum can be simplified further by realising
\begin{align}
    \sum_{l=1}^N\frac{y_l^{2N}}{\prod_{k\neq l}(y_l^2-y_k^2)}=\frac{\det[y_b^{2c-2},y_b^{2N}]_{\substack{b=1,\ldots,N\\c=1,\ldots,N-1}}}{\Delta_N(y^2)}
\end{align}
following from a Laplace expansion in the last column. Furthermore, it is
\begin{align}
\begin{split}
    \frac{\det[y_b^{2c-2},y_b^{2N}]_{\substack{b=1,\ldots,N\\c=1,\ldots,N-1}}}{\Delta_N(y^2)}&=\oint\frac{{\rm d}z}{2\pi z^N}\frac{\det[(-z)^{b-1},y_c^{2b-2}]_{\substack{b=1,\ldots,N+1\\c=1,\ldots,N}}}{\Delta_N(y^2)}\\
    &=\oint\frac{{\rm d}z}{2\pi z^N}\frac{\Delta_{N+1}(-z,y^2)}{\Delta_N(y^2)}=\oint\frac{{\rm d}z}{2\pi z^N}\prod_{l=1}^N(y_l^2+z)\\
    &=\sum_{l=1}^Ny_l^2,
\end{split}
\end{align}
where we transposed the matrix in the determinant in the first step due to formatting. The contour integral is simply deleting the first column and the $N$'th row in the determinant.

Summarizing, the von Neumann entropy of $\rho^{(A)}$ at finite system size $A$ takes the simple form
\begin{align}
\begin{split}\label{A.entropy.fixed.a}
\langle S_A\rangle&=N_A\log(2)-\frac{N_A(2N_A-1)}{4N^2}\sum_{l=1}^Ny_l^{2}+o\left(\frac{1}{N^2}\sum_{l=1}^Ny_l^{2}\right).
\end{split}
\end{align}
This means that the leading order is given by the maximally possible entropy as expected.

When system $A$ is finite it means system $B$ is infinitely large because of $N_B=N-N_A$. We employ~\eqref{av.entropy.f} for $m=N-N_A$ and start with
\begin{align}
\begin{split}
\langle S_B\rangle_y&=(N-N_A)\log(2)+N_A\int_{0}^1{\rm d}\chi\int_1^\infty\frac{{\rm d}t}{t}\oint_{|\eta|=1/2}\frac{{\rm d}\eta}{2\pi\ii }\frac{\eta(\sqrt{t}-1)}{(1-\eta)(1-\chi)(\chi^2-\eta^2) }\\
&\quad\,\,\times\left(\frac{1-\chi}{1-\eta}\right)^{2N_A}\left(\frac{\chi}{\eta}\right)^{2N-2N_A}\left[1-\prod_{k=1}^N\frac{1-\eta^2y_k^2/t}{1-\chi^2y_k^2/t}\right].
\end{split}
\end{align}
Since $N_A$ is fixed and the integrand behave like $1/\eta^2$ for $|\eta|\to\infty$ we can also close the contour around the other pole  at $\eta=1$. Thus, we shift and rescale $\eta\to 1-\eta/(2N)$. Similarly, the maximum in $\chi$ is attained close to $\chi=1$ due to the factor $\chi^{2N-2N_A}$ which pushes the maximum as far away from the origin as possible. Therefore, we additionally reflect and rescale $\chi\to 1-\chi/(2N)$. This leads to
\begin{align}
\begin{split}
\langle S_B\rangle_y&=(N-N_A)\log(2)-NN_A\int_{0}^{2N}{\rm d}\chi\int_1^\infty\frac{{\rm d}t}{t}\oint_{|\eta|=1}\frac{{\rm d}\eta}{2\pi\ii }\frac{(1-\eta/(2N))(\sqrt{t}-1)}{\eta\chi(\chi-\eta)(1-[\chi+\eta]/(4N)) }\\
&\quad\,\,\times\left(\frac{\chi}{\eta}\right)^{2N_A}\left(\frac{1-\chi/(2N)}{1-\eta/(2N)}\right)^{2N-2N_A}\left[1-\prod_{k=1}^N\frac{1-(1-\eta/(2N))^2y_k^2/t}{1-(1-\chi/(2N))^2y_k^2/t}\right].
\end{split}
\end{align}
When we expand the integrand up to order $\mathcal{O}(1/N)$, we find the volume order as well as the constant order of the von Neumann entropy, 
\begin{align}
\begin{split}
\langle S_B\rangle_y&=N\mathcal{S}_1+\mathcal{S}_0+\mathcal{O}\left(\frac{1}{N}\right)
\end{split}
\end{align}
with
\begin{align}
\begin{split}\label{B.entropy.fixed.c}
\mathcal{S}_1&=\log(2)-N_A\int_{0}^{\infty}{\rm d}\chi\int_1^\infty\frac{{\rm d}t}{t}\oint_{|\eta|=1}\frac{{\rm d}\eta}{2\pi\ii }\frac{(\sqrt{t}-1)}{\eta\chi(\chi-\eta) }\left(\frac{\chi}{\eta}\right)^{2N_A}e^{\eta-\chi}\\
&\quad\,\,\times \left[1-\exp\left(\frac{1}{N}\sum_{k=1}^N\frac{y_k^2}{t-y_k^2}(\eta-\chi)\right)\right],\\
\mathcal{S}_0&=-N_A\log(2)-\frac{N_A}{4}\int_{0}^{\infty}{\rm d}\chi\int_1^\infty\frac{{\rm d}t}{t}\oint_{|\eta|=1}\frac{{\rm d}\eta}{2\pi\ii }\frac{(\sqrt{t}-1)}{\eta\chi }\left(\frac{\chi}{\eta}\right)^{2N_A}e^{\eta-\chi}\\
&\hspace*{-0.7cm}\,\times \left[1-\chi-\eta-\left(1-\chi-\eta+\frac{1}{N}\sum_{l=1}^N\frac{y_l^2(t+y_l^2)}{(t-y_l^2)^2}(\chi+\eta)\right)\exp\left(\frac{1}{N}\sum_{k=1}^N\frac{y_k^2}{t-y_k^2}(\eta-\chi)\right)\right].
\end{split}
\end{align}

The volume term can be explicitly computed by introducing an auxiliary integral that takes care of the term $1/(\chi-\eta)$, \ie
\begin{align}
\begin{split}
\frac{\exp[a(\eta-\chi)]-\exp\left(b(\eta-\chi)\right)}{\chi-\eta } =\int_a^bdu\,\exp[u(\eta-\chi)].
\end{split}
\end{align}
Since $a=1$ and $b=1+N^{-1}\sum_{k=1}^Ny_k^2/(t-y_k^2)$ are both positive we can carry out the integral over $\chi$ and $\eta$ first to obtain
\begin{align}
\begin{split}
\mathcal{S}_1&=\log(2)-\frac{1}{2}\int_1^\infty\frac{{\rm d}t}{t}\int_a^bdu(\sqrt{t}-1)=\log(2)-\frac{1}{2N}\int_1^\infty\frac{{\rm d}t}{t}(\sqrt{t}-1)\sum_{k=1}^N\frac{y_k^2}{t-y_k^2}.
\end{split}
\end{align}
The remaining integral can be carried out and yields
\begin{align}
\begin{split}
\mathcal{S}_1&=\log(2)-\frac{1}{N}\sum_{k=1}^N\left[y_k{\rm arctanh}(y_k)+\frac{1}{2}\log(1-y_k^2)\right]=\frac{1}{N}\sum_{k=1}^Ns(y_k).
\end{split}
\end{align}

The constant order can be integrated out directly,
\begin{align}
\begin{split}\label{B.entropy.fixed.d}
\mathcal{S}_0&=-N_A\log(2)-\frac{N_A}{4}\int_1^\infty\frac{{\rm d}t}{t}(\sqrt{t}-1)\biggl[\frac{1}{2N_A}-2\\
&\quad\,-\left(\frac{1}{2N_A}-2\frac{1-N^{-1}\sum_{k=1}^Ny_k^2(t+y_k^2)/(t-y_k^2)^2}{1+N^{-1}\sum_{k=1}^Ny_k^2/(t-y_k^2)}\right)\biggl]\\
&=-N_A\log(2)+N_A\int_1^\infty{\rm d}t(\sqrt{t}-1)\frac{N^{-1}\sum_{k=1}^Ny_k^2/(t-y_k^2)^2}{1+N^{-1}\sum_{k=1}^Ny_k^2/(t-y_k^2)}.
\end{split}
\end{align}
The integral against $t$ is in the general case non-trivial. However, when we plug in the maximally degenerate case with $y_1=\ldots=y_k=y$ we find $\mathcal{S}_0=N_A\log(2)-2N_As(y)$ as desired considering the relation~\eqref{ent.rel.max} and the results~\eqref{A.entropy.fixed.a} and~\eqref{B.entropy.fixed.c} in this case. 

In summary, the expected von Neuman entropy of system $B$ in the case of a fixed size of system $A$ is
\begin{align}
\begin{split}\label{B.entropy.fixed.e}
\langle S_B\rangle_y&=\sum_{k=1}^Ns(y_k)-N_A\log(2)+N_A\int_1^\infty{\rm d}t(\sqrt{t}-1)\frac{N^{-1}\sum_{k=1}^Ny_k^2/(t-y_k^2)^2}{1+N^{-1}\sum_{k=1}^Ny_k^2/(t-y_k^2)}+\mathcal{O}\left(\frac{1}{N}\right).
\end{split}
\end{align}
Hence, the mean of the mutual information~\eqref{mutual.dens} is
\begin{align}\label{mut.finite.result}
    \langle I_{AB}\rangle=N_A\int_1^\infty{\rm d}t(\sqrt{t}-1)\frac{N^{-1}\sum_{k=1}^Ny_k^2/(t-y_k^2)^2}{1+N^{-1}\sum_{k=1}^Ny_k^2/(t-y_k^2)}+\mathcal{O}\left(\frac{1}{N}\right),
\end{align}
where we employed~\eqref{A.entropy.fixed.a} and~\eqref{B.entropy.fixed.e}.
This means that in the present case it is of order one where most of the entropy comes form the larger subsystem $B$.

\begin{figure}[t]
    \begin{tikzpicture}
    \draw (0,0) node[inner sep=0pt]{\includegraphics[width=\linewidth]{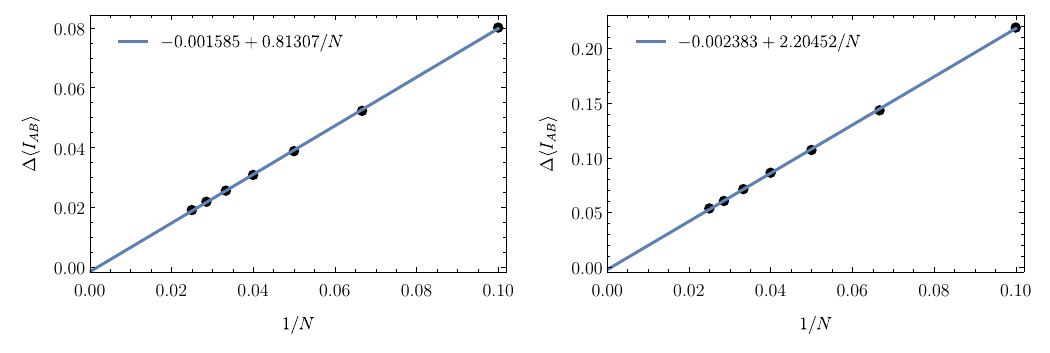}};
    \draw (-7.35,2) node[scale=.7]{\textbf{(a)}} (.4,2) node[scale=.7]{\textbf{(b)}};
    \end{tikzpicture}
    \caption{
    Error scaling of the mutual information for constant subsystem size $N_A=2$ for a maximally degenerate system with $y_1=\ldots=y_N=0.5$ (plot \textbf{(a)}, difference between the exact finite $N$ expression~\eqref{max.deg.rel} and its asymptotic~\eqref{mut.finite.result.degenerate}) and a system with evenly spaced $y_j=(j-1)/(N-1)$ for $j=1,\dots,N$ (plot \textbf{(b)}, difference between Monte-Carlo simulation with $10^7$ samples and~\eqref{mut.finite.result}).
    \label{fig:test-constant-subsystem}}
\end{figure}

The error scaling is shown in Fig.~\ref{fig:test-constant-subsystem}, where we study two different systems. The first one is a maximally degenerate system comparing the exact finite $N$ expression based on~\eqref{max.deg.rel} and its asymptotic~\eqref{mut.finite.result.degenerate}. The second one is a system for evenly spaced $y_j$ for which we take the difference of~\eqref{mut.finite.result} and a Monte-Carlo simulation. In both cases, the error clearly scales as $1/N$.

For a maximally degenerate quantum state of the total system with $y_1=\dots=y_N=y$, we can evaluate the integral in~\eqref{mut.finite.result} to find
\begin{align}\label{mut.finite.result.degenerate}
    \langle I_{AB}\rangle=N_A\left(\log (1-y)+\log (y+1)+2 y \tanh ^{-1}(y)\right)+\mathcal{O}\left(\frac{1}{N}\right)\,.
\end{align}
In the limit of a globally pure state, this simplifies to $\braket{I_{AB}}=2N_A\log(2)+o(1)$, which is expected for a globally pure state with a small subsystem, \ie $\lim_{N\to\infty}N_A/N=0$, as it is exactly twice the maximal entanglement entropy of a subsystem of size $N_A$.

\subsection{Subsystem \texorpdfstring{$A$}{A} of size \texorpdfstring{$N_A=f N$}{NA=fN}}\label{sec:large.subsystem}
Next, we consider the situation where the relative size $f=N_A/N$ of system $A$ approaches a number in $(0,1)$. This implies also that the number of modes in $B$ also scales with $N$, \ie $N_B/N=1-f\in(0,1)$. Thence, it is sufficient to consider the limit of~\eqref{av.entropy.b} for $f=N_A/N\in(0,1)$ as the reflection $f\leftrightarrow1-f$ of the result covers the complementary subsystem, as well.

This time we start from~\eqref{av.entropy.f}, \ie
\begin{align}
\begin{split}
\langle S(\rho^{( f  N)})\rangle_y={}& f  N\log(2)+(1- f )N\int_{0}^1{\rm d}\chi\int_1^\infty\frac{{\rm d}t}{t}\oint_{|\eta|=1/2}\frac{{\rm d}\eta}{2\pi\ii }\frac{\eta(\sqrt{t}-1)}{(1-\eta)(1-\chi)(\chi^2-\eta^2) }\\
&\times\left(\frac{1-\chi}{1-\eta}\right)^{2N(1- f )}\left(\frac{\chi}{\eta}\right)^{2N f }\left[1-\prod_{k=1}^N\frac{1-\eta^2y_k^2/t}{1-\chi^2y_k^2/t}\right].
\end{split}
\end{align}
The tricky thing is that there are two terms in the integral which need to be dealt with accordingly. The idea is to write them both as the following integral
\begin{align}
    1-\prod_{k=1}^N\frac{1-\eta^2y_k^2/t}{1-\chi^2y_k^2/t}=-\sum_{l=1}^N\log\left(\frac{1-\eta^2y_l^2/t}{1-\chi^2y_l^2/t}\right)\int_0^1{\rm d}u\prod_{k=1}^N\frac{(1-\eta^2y_k^2/t)^u}{(1-\chi^2y_k^2/t)^u}.
\end{align}
There is no issue with branch cuts due to $0\leq|\eta^2y_k^2/t|,\chi^2y_k^2/t<1$.
We may interchange all four integrals such that we carry out first the integrals over $\chi$ and $\eta$ for which we perform the saddle point analysis.

The saddle point equations are
\begin{align}\label{saddle.2}
    0=\frac{2N f }{\chi}-\frac{2N(1- f )}{1-\chi}+u\sum_{k=1}^N\frac{2\chi y_k^2}{t-\chi^2 y_k^2},\quad  0=\frac{2N f }{\eta}-\frac{2N(1- f )}{1-\eta} +u\sum_{k=1}^N\frac{2\eta y_k^2}{t-\eta^2 y_k^2}.
\end{align}
The derivatives in $t$ as well as $u$ only tell us that $\chi$ and $\eta$ must take the same saddle points but they do not indicate any distinct points for $t$ and $u$.
These equations have a unique solution when considering the restriction  $0<|\eta|,\chi<1$. Surely, they are equivalent to polynomial equations of degree $2N+1$ with real coefficients when assuming that all $y_k$ are pairwise distinct and none is equal to $1$\ \footnote{If this is not the case the degree of the polynomial decreases but the following argumentation works out along the same lines with only minor modifications. The final results, however, are unaffected.}.   When ordering $y_1<y_2<\ldots<y_N<1$, one can readily find $N-1$ real solutions, one in each interval $(\sqrt{t}/y_{k+1},\sqrt{t}/y_k)$ for all $k=1,\ldots,N-1$, since the rational function on the right hand sides in~\eqref{saddle.2} diverges to $-\infty$ when $\chi\searrow t/y_{k+1}$ and to $+\infty$ when $\chi\nearrow t/y_{k}$. Another  $N-1$ solutions are on the negative real axis due to the same reason. There must be also a zero of the rational function in~\eqref{saddle.2} in the interval $(0,1)$ as it diverges to $+\infty$ when $\chi\searrow0$ and $-\infty$ when $\chi\nearrow1$. The last remaining two zeros  lie in $(\sqrt{t}/y_1^2,+\infty)$ and $(-\infty,-\sqrt{t}/y_1^2)$ as the leading order term of the rational function is $2N(1-u)/\chi$ when $u\in[0,1)$ while it diverges to $-\infty$ when $\chi\searrow t/y_{1}$ and to  $+\infty$ when $\chi\nearrow -t/y_{1}$, respectively. 

A closer analysis shows that the solution in $(0,1)$ lies even in $[f,1)$. We can give the solution of~\eqref{saddle.2} in $(0,1)$ an analytical expression in terms of an integral over a Heaviside step function, namely
\begin{align}\label{sol.2}
    \eta=\chi=h( f ,y,t,u)=\int_0^1 {\rm d}\lambda\ \Theta\left(\frac{ f }{\lambda}-\frac{1- f }{1-\lambda}+\frac{u}{N}\sum_{k=1}^N\frac{\lambda y_k^2}{t-\lambda^2 y_k^2}\right)\in[f,1).
\end{align}
For $u=0$ or $f\to\infty$ it is evidently $h( f ,y,t,0)=h( f ,y,\infty,u)=f$.

Let us note that the moving simple pole at $\chi=\eta$ is removed by the terms $\log\left([1-\eta^2y_l^2/t]/[1-\chi^2y_l^2/t]\right)$. Therefore, we expand as follows
\begin{align}
    \chi=h( f ,y,t,u)+\delta\chi/\sqrt{N} \qquad{\rm and}\qquad \eta=h( f ,y,t,u)+\ii\delta\eta/\sqrt{N},
\end{align}
which reflects how the parameters $\chi$ and $\eta$ traverse through $h( f ,y,t,u)$. It is for the non-exponential prefactors
\begin{align}
\begin{split}
  &\frac{\eta(\sqrt{t}-1)}{t(1-\eta)(1-\chi)(\chi^2-\eta^2) } \sum_{l=1}^N\log\left(\frac{1-\eta^2y_l^2/t}{1-\chi^2y_l^2/t}\right)\\
  ={}&NH_{0,0}( f ,y,t,u)+\sqrt{N}\left[H_{1,0}( f ,y,t,u)\delta\chi+\ii H_{0,1}( f ,y,t,u)\delta\eta\right]\\
  &+\left[H_{2,0}( f ,y,t,u)\delta\chi^2+\ii H_{1,1}( f ,y,t,u)\delta\chi\delta\eta-H_{0,2}( f ,y,t,u)\delta\eta^2\right]+\mathcal{O}\left(\frac{1}{\sqrt{N}}\right) 
\end{split}
\end{align}
with
\begin{align}
        H_{0,0}( f ,y,t,u)={}&\frac{\sqrt{t}-1}{t(1-h( f ,y,t,u))^2 } \frac{1}{N}\sum_{l=1}^N\frac{h( f ,y,t,u)y_l^2}{t-h^2( f ,y,t,u)y_l^2},\label{H00}\\
        H_{1,0}( f ,y,t,u)={}&\frac{\sqrt{t}-1}{t(1-h( f ,y,t,u))^3 } \frac{1}{N}\sum_{l=1}^N\frac{h( f ,y,t,u)y_l^2}{[t-h^2( f ,y,t,u)y_l^2]^2}\nonumber\\
        &\times(t+y_l^2h( f ,y,t,u)-2y_l^2h^2( f ,y,t,u)),\\
        H_{0,1}( f ,y,t,u)={}&\frac{\sqrt{t}-1}{t(1-h( f ,y,t,u))^3 } \frac{1}{N}\sum_{l=1}^N\frac{y_l^2(t-h^3( f ,y,t,u)y_l^2)}{[t-h^2( f ,y,t,u)y_l^2]^2},\\
        H_{2,0}( f ,y,t,u)={}&\frac{\sqrt{t}-1}{6t(1-h( f ,y,t,u))^4 } \frac{1}{N}\sum_{l=1}^N\frac{h( f ,y,t,u)y_l^2}{[t-h^2( f ,y,t,u)y_l^2]^3}\biggl[6t^2+3ty_l^2\nonumber\\
        &\hspace*{-1cm}+(5y_l^4-15ty_l^2)h^2( f ,y,t,u)-16y_l^4h^3( f ,y,t,u)+17y_l^4h^4( f ,y,t,u)\biggl],\\
        H_{1,1}( f ,y,t,u)={}&\frac{\sqrt{t}-1}{3t(1-h( f ,y,t,u))^4 } \frac{1}{N}\sum_{l=1}^N\frac{y_l^2}{[t-h^2( f ,y,t,u)y_l^2]^3}\nonumber\\
        &\times\biggl[3t^2+3ty_l^2h( f ,y,t,u)-6ty_l^2h^2( f ,y,t,u)+(y_l^4-3ty_l^2)\nonumber\\
        &\times h^3( f ,y,t,u)-5y_l^4h^4( f ,y,t,u)+7y_l^4h^5( f ,y,t,u)\biggl],\\
        H_{0,2}( f ,y,t,u)={}&\frac{\sqrt{t}-1}{6t(1-h( f ,y,t,u))^4 } \frac{1}{N}\sum_{l=1}^N\frac{y_l^2}{[t-h^2( f ,y,t,u)y_l^2]^3}\nonumber\\
        &\times\biggl[6t^2+9ty_l^2h( f ,y,t,u)-24ty_l^2h^2( f ,y,t,u)+(3ty_l^2-y_l^4)\nonumber\\
        &\times h^3( f ,y,t,u)+2y_l^4h^4( f ,y,t,u)+5y_l^4h^5( f ,y,t,u)\biggl].
\end{align} 
The Jacobian of this substitution yields the factor $\ii/N$. The exponential factor has the expansion
\begin{align}
\begin{split}
   &\hspace{-1.5cm} \left(\frac{1-\chi}{1-\eta}\right)^{2N(1- f )}\left(\frac{\chi}{\eta}\right)^{2N f }\prod_{k=1}^N\frac{(1-\eta^2y_k^2/t)^u}{(1-\chi^2y_k^2/t)^u}\\
   ={}&\exp\left[-L_2( f ,y,t,u)(\delta\chi^2+\delta\eta^2)\right]\biggl[1-L_3( f ,y,t,u)\frac{\delta\chi^3+\ii\delta\eta^3}{\sqrt{N}}\\
   &+L_3^2( f ,y,t,u)\frac{(\delta\chi^3+\ii\delta\eta^3)^2}{2N}-L_4( f ,y,t,u)\frac{\delta\chi^4-\delta\eta^4}{N}+\mathcal{O}\left(\frac{1}{N^{3/2}}\right)\biggl]
\end{split}
\end{align}
with
\begin{align}
    L_2( f ,y,t,u)={}&\frac{ f }{h^2( f ,y,t,u)}+\frac{1- f }{(1-h( f ,y,t,u))^2}-\frac{u}{N}\sum_{k=1}^N\frac{y_k^2[t +y_k^2h^2( f ,y,t,u)]}{(t-h^2( f ,y,t,u)y_k^2)^2},\\
    L_3( f ,y,t,u)={}&-\frac{2}{3}\biggl[\frac{ f }{h^3( f ,y,t,u)}-\frac{1- f }{(1-h( f ,y,t,u))^3}\nonumber\\
    &+\frac{u}{N}\sum_{k=1}^N\frac{y_k^4h( f ,y,t,u)}{(t-h^2( f ,y,t,u)y_k^2)^3}[3t +y_k^2h^2( f ,y,t,u)]\biggl],\\
    L_4( f ,y,t,u)={}&\frac{1}{2}\biggl[\frac{ f }{h^4( f ,y,t,u)}+\frac{1- f }{(1-h( f ,y,t,u))^4}-\frac{u}{N}\sum_{k=1}^N\frac{y_k^4}{(t-h^2( f ,y,t,u)y_k^2)^4}\nonumber\\
    &\times[t^2 +6ty_k^2h^2( f ,y,t,u)+y_k^4h^4( f ,y,t,u)]\biggl].
\end{align}
In spite of the amount of terms to obtain the leading and next to leading term for the von Neumann entropy, the integral  over $\delta\chi$ and $\delta\eta$ simplifies many terms because odd moments of a centred Gaussian vanish and the Gaussians in $\delta\chi$ and $\delta\eta$ have the same variance. Therefore, we arrive at the following approximation
\begin{align}\label{entropy.limit}
\begin{split}
\langle S(\rho^{( f  N)})\rangle_y={}& N\left[f  \log(2)-\frac{1- f}{2}\int_1^\infty {\rm d}t\int_0^1{\rm d}u\frac{H_{0,0}( f ,y,t,u)}{L_2( f ,y,t,u)}\right]\\
&-\frac{1- f}{2}\int_1^\infty {\rm d}t\int_0^1{\rm d}u\biggl[\frac{H_{2,0}( f ,y,t,u)-H_{0,2}( f ,y,t,u)}{2L_2^2( f ,y,t,u)}\\
&-\frac{3L_3( f ,y,t,u)}{4L_2^3( f ,y,t,u)}(H_{1,0}( f ,y,t,u)-H_{0,1}( f ,y,t,u))\biggl]+\mathcal{O}\left(\frac{1}{N}\right).
\end{split}
\end{align}

\begin{figure}[t]
    \begin{tikzpicture}
    \draw (0,0) node[inner sep=0pt]{\includegraphics[width=\linewidth]{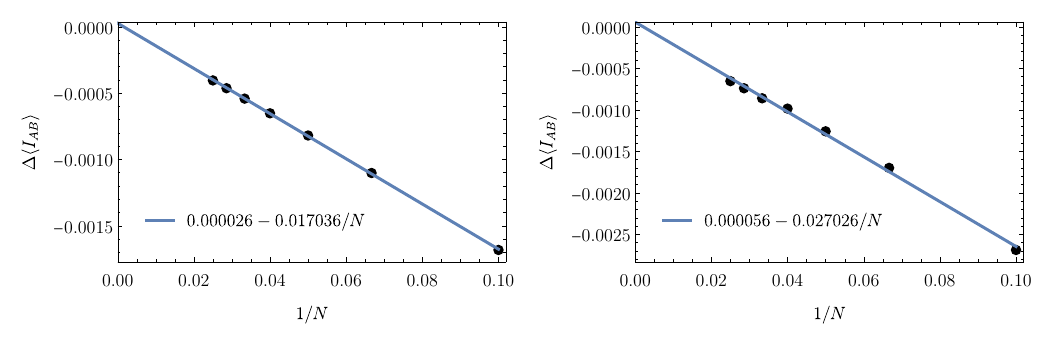}};
    \draw (-7.35,2) node[scale=.7]{\textbf{(a)}} (.4,2) node[scale=.7]{\textbf{(b)}};
    \end{tikzpicture}
    \caption{
    Error scaling of the mutual information for constant subsystem fraction $f=N_A/N=2/5$. Test of formula~\eqref{mutual.limit}.  We consider a maximally degenerate system with $y_1=\ldots=y_N=0.5$ (plot \textbf{(a)}, difference between~\eqref{mutual.limit} and the numerical evaluation of the exact expression~\eqref{max.deg.rel}) and a system with evenly spaced $y_j=(j-1)/(N-1)$ for $j=1,\dots,N$  (plot \textbf{(b)}, difference between~\eqref{mutual.limit} and a Monte-Carlo simulation with $10^7$ samples).
    \label{fig:test-constant-subsystem-fraction}}
\end{figure}

We can simplify this expression further by exploiting that the saddle point equation holds for all $u$ and $t$. In particular, its total derivative in $u$ gives the relation
\begin{equation}
    L_2( f ,y,t,u)\partial_uh( f ,y,t,u)=\frac{1}{N}\sum_{k=1}^N\frac{h( f ,y,t,u)y_k^2}{t-h^2( f ,y,t,u)y_k^2}.
\end{equation}
Employing~\eqref{H00} we are allowed to carry out the $u$-integral in the volume law term and find
\begin{align}
\begin{split}
&f  \log(2)-\frac{1- f}{2}\int_1^\infty {\rm d}t\int_0^1{\rm d}u\frac{H_{0,0}( f ,y,t,u)}{L_2( f ,y,t,u)}\\
&=f  \log(2)-\frac{1- f}{2}\int_1^\infty {\rm d}t\int_f^{h( f ,y,t,1)}{\rm d}h\,\frac{\sqrt{t}-1}{t(1-h)^2}\\
&=f  \log(2)-\frac{1}{2}\int_1^\infty {\rm d}t\,\frac{(\sqrt{t}-1)(h( f ,y,t,1)-f)}{t(1-h( f ,y,t,1))}
\end{split}
\end{align}
The following relations also hold true
\begin{equation}
\begin{split}
    H_{1,0}( f ,y,t,u)-H_{0,1}( f ,y,t,u))&=-\frac{H_{0,0}( f ,y,t,u)}{h( f ,y,t,u)},\\
    H_{2,0}( f ,y,t,u)-H_{0,2}( f ,y,t,u)&=-\frac{H_{1,0}( f ,y,t,u)}{h( f ,y,t,u)},
\end{split}
\end{equation}
which simplifies the von Neumann entropy even further,
\begin{align}\label{entropy.limit.b}
\begin{split}
\langle S(\rho^{( f  N)})\rangle_y=& N\left[f  \log(2)-\frac{1}{2}\int_1^\infty {\rm d}t\,\frac{(\sqrt{t}-1)(h( f ,y,t,1)-f)}{t(1-h( f ,y,t,1))}\right]\\
&+\frac{1- f}{2}\int_1^\infty {\rm d}t\int_0^1\frac{{\rm d}u}{h( f ,y,t,u)}\biggl[\frac{H_{1,0}( f ,y,t,u)}{2L_2^2( f ,y,t,u)}\\
&-\frac{3L_3( f ,y,t,u)}{4L_2^3( f ,y,t,u)}H_{0,0}( f ,y,t,u)\biggl]+\mathcal{O}\left(\frac{1}{N}\right).
\end{split}
\end{align}
This result can be put together for the mutual information which is
\begin{equation}\label{mutual.limit}
    \langle I_{AB}\rangle_y=\langle S(\rho^{( f  N)})\rangle_y+\langle S(\rho^{( (1-f)  N)})\rangle_y-\sum_{k=1}^Ns(y_k).
\end{equation}
We show the error scaling in Fig.~\ref{fig:test-constant-subsystem-fraction}, where we compare~\eqref{mutual.limit} with the analytics for a maximally degenerate system as well as with the results of a Monte-Carlo simulation for evenly spaced spectrum of $y_j$. In both cases, we clearly see that the error scales as $1/N$.

\subsection{Leading order statistics at maximal degeneracy}\label{sec:case-study}

We would like to come back to the case of maximal degeneracy with a singular value $y\in[0,1]$ of the complex structure $J$. We recall that the case $y=1$ corresponds to a pure Gaussian quantum state, as considered in~\cite{bianchi2021page}, while $y=0$ is the maximally mixed quantum state.

We could use the results~\eqref{entropy.limit} and~\eqref{mutual.limit} for $y_1=\ldots=y_N=y$. The saddle point solution~\eqref{sol.2} would then follow from a cubic polynomial equation. This renders the integrals in~\eqref{entropy.limit} cumbersome as they are designed for the general setting. As mentioned in Sec.~\ref{sec:example.2}, there is a  more direct way to compute the von Neumann entropy and the mutual information in the case of maximal degeneracy, namely via its simple relation to reduced density matrices of fermionic Gaussian pure quantum states.

We make use of~\eqref{max.deg.rel} and employ the macroscopic level density
\begin{align}
    \varrho_{\rm max}^{(f)}(x)=\lim_{N\to\infty}\varrho^{(N,N_B)}(x)&=\frac{2}{y\pi(1-\sqrt{1-a^2})}\frac{\sqrt{a^2-x^2/y^2}}{1-x^2/y^2},\label{eq:level-density}
\end{align}
where $a=\sqrt{4f(1-f)}$ and its support is $x\in[0,ay]$. This reproduces the level density for a pure state in the limit $y=1$. Let us highlight again that the level density~\eqref{eq:level-density} only captures the continuous part of the spectrum, which is symmetric under $f\to 1-f$. For $f\leq 1/2$, the spectrum consists only of a density, while for $f>\frac{1}{2}$ there will be $N_A-N_B=N(2f-1)$ singular values at $y$. Without loss of generality we can restrict ourselves to $f\leq 1/2$ because the mutual information is symmetric in systems $A$ and $B$ and we can exploit~\eqref{ent.rel.max} and~\eqref{ent.rel.max}.

Therefore, we compute only $\braket{S_A}$ for a fixed subsystem fraction $f=N_A/N\leq 1/2$ and have at leading order
\begin{align}
   \lim_{N\to\infty}\frac{\braket{S_A}}{N}&=f\int^{ay}_{0}\varrho_{\rm max}^{(f)}(x)s(x){\rm d}x=f\left(\log(2)-\frac{1}{2}\int^{ay}_{-ay}\varrho_{\rm max}^{(f)}(x)(1+x)\log(1+x){\rm d}x\right),
\end{align}
where we have used the symmetry $\varrho_{\rm max}^{(f)}(x)=\varrho_{\rm max}^{(f)}(-x)$. Plugging in~\eqref{eq:level-density} and rescaling $x=ayz$, we arrive at
\begin{align}
    \lim_{N\to\infty}\frac{\braket{S_A}}{N}&=f\left(\log(2)-\frac{a^2}{\pi(1-\sqrt{1-a^2})}\int^{1}_{-1}\frac{\sqrt{1-z^2}}{1-a^2z^2}(1+ayz)\log(1+ayz){\rm d}z\right)\nonumber\\
    &=f\biggl[\log(2)+{\log(\beta)}-{\frac{ay}{\pi(1-\sqrt{1-a^2})}}\underbrace{\int^{1}_{-1}\frac{\sqrt{1-z^2}}{\alpha^2-z^2}(\beta+z)\log(\beta+z){\rm d}z}_{=:I_0}\biggl],\label{SA.splitting}
\end{align}
for which we have introduced the integral $I_0$ with the parameters $\alpha=1/a$ and $\beta=1/(ay)$ that satisfy $\beta\geq \alpha\geq 1$. 

The problem of the integral $I_0$ is the combined logarithmic and square root branch cuts. It can be computed by splitting it into two pieces $I_0=I_1+I_2$ with
\begin{align}
    I_1&=\int^{1}_{-1}\sqrt{1-z^2}\left(\frac{1}{\alpha^2-z^2}-\frac{1}{1-z^2}\right)(\beta+z)\log(\beta+z){\rm d}z,\\
    I_2&=\int^{1}_{-1}\frac{\beta+z}{\sqrt{1-z^2}}\log(\beta+z){\rm d}z.
\end{align}
The integral $I_2$ can be evaluated classically. For instance, it is
\begin{equation}
    \int^{1}_{-1}\frac{z}{\sqrt{1-z^2}}\log(\beta+z){\rm d}z=-\int^{1}_{-1}\log(\beta+z)\partial_z\sqrt{1-z^2}{\rm d}z=\int^{1}_{-1}\frac{\sqrt{1-z^2}{\rm d}z}{\beta+z},
\end{equation}
which is, up to a normalisation factor of $1/\pi$, the Green function of Wigner's semi-circle~\cite[Eq.~(1.132)]{log-gases-peter}, \ie
\begin{equation}
    \int^{1}_{-1}\frac{z}{\sqrt{1-z^2}}\log(\beta+z){\rm d}z=\pi  \left[\beta-\sqrt{\beta^2-1}\right].
\end{equation}
In contrast, the first part of $I_2$ can be understood as $\beta$ times the antiderivative of
\begin{equation}\label{intermediate.result}
    \int^{1}_{-1}\frac{1}{\sqrt{1-z^2}}\frac{{\rm d}z}{\beta+z}=\frac{\pi}{\sqrt{\beta^2-1}},
\end{equation}
which is the Green function of the arcsine law. It can be computed by only integrating the even part $\beta/[\sqrt{1-z^2}(\beta^2-z^2)]$ of the integrand and substituting $z=\cos[\varphi/2]$ with $\varphi\in[0,2\pi]$. The resulting integral can be computed with the help of residue theorem when identifying the complex variable $z'=e^{\ii\varphi}$. 

The antiderivative in $\beta$ of the intermediate result~\eqref{intermediate.result} with the proper asymptotic behaviour for $\beta\to\infty$ is $\pi\log[(\sqrt{\beta^2-1}+\beta)/2]$. Thus, we arrive at
\begin{equation}\label{I2.result}
\begin{split}
    I_2={}&\pi  \left[\beta \log \left(\frac{\sqrt{\beta^2-1}+\beta}{2}\right)+\beta-\sqrt{\beta^2-1}\right]\\
    ={}&\frac{\pi}{a y}\left[\log \left(\frac{\sqrt{1-a^2y^2}+1}{2ay}\right)+1-\sqrt{1-a^2y^2}\right].
\end{split}
\end{equation}

For the remaining integral $I_1$ we need more involved complex analysis methods. The essential part of our restructuring is that this integral has an integrable branch point at $z=\infty$ which allows us certain contour deformations. In the first step, we rewrite
\begin{align}
    I_1=\int_\gamma g(z){\rm d}z\quad\text{with}\quad g(z)=\frac{\alpha^2-1}{2\ii}\frac{(\beta+z)\log(\beta+z)}{(\alpha^2-z^2)\sqrt{z-1}\sqrt{z+1}},\label{eq:I1-pre}
\end{align}
where the infinitesimal surrounding of the square root branch cut on $[-1,1]$ of $1/(\sqrt{z-1}\sqrt{z+1})$ gives the monodromy $-2\ii/\sqrt{1-z^2}$ along the cut, see Fig.~\ref{fig:contour}. We would like to point out that we use the principal branch of the complex square root along the negative real axis. Additionally, we underline that the behaviour at $z\to\infty$ of the integrand is $g(z)=\mathcal{O}(\log(z)/z^2)$ highlighting the integrability at infinity.

\begin{figure}
    \centering
    \begin{tikzpicture}
        \draw[thick,->] (-4,0) -- (4,0) node[below]{$\mathrm{Re}(z)$};
        \draw[thick,->] (0,-1) -- (0,2) node[right]{$\mathrm{Im}(z)$};

        \draw[blue,very thick] (-1,0) -- (1,0);
        \fill[blue] (-1,0) node[yshift=-5mm]{$-1$}  circle(.08) (1,0) node[yshift=-5mm]{$1$} circle(.08);

        \draw[thick,red] ([shift=(90:.2cm)]1,0) arc (90:-90:.2cm) -- ([shift=(-90:.2cm)]-1,0) arc (-90:-270:.2cm) -- cycle;
        \draw[thick,purple] (-4,.2) -- ([shift=(90:.2cm)]-3,0) arc (90:-90:.2cm) -- (-4,-.2);
        \draw[thick,dashed] (-5,.2) -- (-4,.2) (-5,-.2) -- (-4,-.2);

        \draw (0,-2) node{$\displaystyle I_1={\color{red}\oint_{\gamma} g(z)dz}={\color{purple}\oint_{\gamma'} g(z)dz}+2\pi\ii(\mathrm{Res}_{\alpha}(g)+\mathrm{Res}_{-\alpha}(g))$};

        \draw[thick,->,red] (0,.2) -- (.5,.2) node[above]{$\gamma$};
        \draw[thick,->,purple] (-3,.2) -- (-3.5,.2) node[above]{$\gamma'$};
        \draw[thick,->] ([shift=(95:.2cm)]2,0) arc (95:455:.2cm);
        \draw[thick,->] ([shift=(95:.2cm)]-2,0) arc (95:455:.2cm);

        \fill (-2,0) node[yshift=-5mm]{$-\alpha$} circle(.08);
        \fill (2,0) node[yshift=-5mm]{$\alpha$}  circle(.08);

        \draw[dgreen,very thick, dashed] (-5,0) -- (-4,0);
        \draw[dgreen,very thick] (-4,0) -- (-3,0);
        \fill[dgreen] (-3,0) node[yshift=-5mm]{$-\beta$} circle(.08);
        
    \end{tikzpicture}
    \caption{The contour deformation of the integral  $I_1$ where we start from $\gamma$, see~\eref{eq:I1-pre}, and end up with $\gamma'$, {\it cf.}, Eq.~\eref{eq:I1}. Thereby we pick up the residues at $\pm\alpha$. }
    \label{fig:contour}
\end{figure}

In order to evaluate~\eqref{eq:I1-pre}, we want to relate the integral surrounding the square root branch cut along the interval $[-1,1]$ with the integral encircling the logarithmic branch cut along $(-\infty,-\beta)$. On the compactified complex plane (Riemann sphere), we can continuously deform the contour $\gamma$ to $\gamma'$, but we will pick up the two residues when crossing the simple poles at $\pm \alpha$. Hence, it is
\begin{align}
    I_1=\oint_{\gamma'}g(z){\rm d}z+2\pi\ii[\mathrm{Res}_{\alpha}(g)+\mathrm{Res}_{-\alpha}(g)],\label{eq:I1}
\end{align}
where the two residues can be evaluated straightforwardly as
\begin{align}\label{residues}
    \mathrm{Res}_{\pm \alpha}(g)&={\frac{\ii}{4}\sqrt{\alpha^2-1}\left(\frac{\beta}{\alpha}\pm1\right)}\log(\beta\pm\alpha).
\end{align}
The remaining integral simplifies tremendously, as it will allow us to get rid of the logarithm via its monodromy $\lim_{\epsilon\to 0}[\log(-t+\ii\epsilon)-\log(-t-\ii\epsilon)]=2\pi \ii$. Thence, we compute
\begin{align}
\begin{split}
    \oint_{\gamma'} g(z){\rm d}z&={-}\lim_{\epsilon\to 0}\int^\infty_{\beta}\big[g(-t+\ii\epsilon)-g(-t-\ii\epsilon)\big]{\rm d}t={\pi(\alpha^2-1)\int^\infty_{\beta}\frac{(\beta-t){\rm d}t}{(\alpha^2-t^2)\sqrt{t^2-1}}.}
\end{split}
\end{align}
This integral can be readily evaluated when substituting $t=\cosh(\vartheta)$,
\begin{align}
\begin{split}
    \oint_{\gamma'} g(z)dz&=\pi(\alpha^2-1)\int^\infty_{{\rm arccosh}(\beta)}\frac{\beta-\cosh(\vartheta)}{\alpha^2-\cosh^2(\vartheta)}{\rm d}\vartheta\\
    &=\pi\frac{\alpha^2-1}{2\alpha}\int^\infty_{{\rm arccosh}(\beta)}\left[\frac{\alpha+\beta}{\alpha+\cosh(\vartheta)}-\frac{\alpha-\beta}{\alpha-\cosh(\vartheta)}\right]{\rm d}\vartheta.
\end{split}
\end{align}
The antiderivative of $1/[\pm\alpha-\cosh(\vartheta)]$ follows from
\begin{equation}
    \partial_\vartheta\left(\frac{2}{\sqrt{\alpha^2-1}}{\rm arctanh}\left[\sqrt{\frac{\alpha\pm1}{\alpha\mp1}}\tanh\left(\frac{\vartheta}{2}\right)\right]\right)=\frac{1}{\pm\alpha-\cosh(\vartheta)},
\end{equation}
and combining this with the relation between the logarithm and ${\rm arctanh}$ we obtain
\begin{align}
\begin{split}\label{contour.result}
    \oint_{\gamma'} g(z)dz={}&\pi\frac{\sqrt{\alpha^2-1}}{2\alpha}\biggl[(\beta+\alpha)\log\left(\frac{(\sqrt{\alpha+1}+\sqrt{\alpha-1})^2(\beta+\alpha)}{(\sqrt{(\alpha+1)(\beta+1)}+\sqrt{(\alpha-1)(\beta-1)})^2}\right)\\
    &{+(\beta-\alpha)\log\left(\frac{(\sqrt{\alpha+1}+\sqrt{\alpha-1})^2(\beta-\alpha)}{(\sqrt{(\alpha+1)(\beta-1)}+\sqrt{(\alpha-1)(\beta+1)})^2}\right)}\biggl].
\end{split}
\end{align}
Putting~\eqref{residues} and~\eqref{contour.result} into~\eqref{eq:I1}, the second integral is
\begin{align}
\begin{split}
    I_1={}&\pi\frac{\sqrt{\alpha^2-1}}{\alpha}\biggl[(\beta+\alpha)\log\left(\frac{\sqrt{\alpha+1}+\sqrt{\alpha-1}}{\sqrt{(\alpha+1)(\beta+1)}+\sqrt{(\alpha-1)(\beta-1)}}\right)\\
    &{+(\beta-\alpha)\log\left(\frac{\sqrt{\alpha+1}+\sqrt{\alpha-1}}{\sqrt{(\alpha+1)(\beta-1)}+\sqrt{(\alpha-1)(\beta+1)}}\right)}\biggl]\\
    ={}&\pi\sqrt{1-a^2}\biggl[\frac{1+y}{ay}\log\left(\frac{(\sqrt{1+a}+\sqrt{1-a})\sqrt{ay}}{\sqrt{(1+a)(1+ay)}+\sqrt{(1-a)(1-ay)}}\right)\\
    &{+\frac{1-y}{ay}\log\left(\frac{(\sqrt{1+a}+\sqrt{1-a})\sqrt{ay}}{\sqrt{(1+a)(1-ay)}+\sqrt{(1-a)(1+ay)}}\right)}\biggl].
\end{split}
\end{align}
This integral and~\eqref{I2.result} can be combined in~\eqref{SA.splitting} for the average entropy of system $A$. Due to the logarithms several terms are canceling so that it is
\begin{equation}
\begin{split}
    \lim_{N\to\infty}\frac{\braket{S_A}}{fN}={}&\log(2)-\frac{1-\sqrt{1-a^2y^2}}{1-\sqrt{1-a^2}}-\frac{\sqrt{1-a^2}}{1-\sqrt{1-a^2}}\log\left[\frac{1+\sqrt{1-a^2}}{2}\right]\\
    &-\frac{\log\left[(1+\sqrt{1-a^2y^2})/2\right]}{1-\sqrt{1-a^2}}+\frac{\sqrt{1-a^2}}{1-\sqrt{1-a^2}}\\
    &\times\biggl[\frac{1+y}{2}\log\left(\frac{1+a^2y+\sqrt{(1-a^2)(1-a^2y^2)}}{2}\right)\\
    &+\frac{1-y}{2}\log\left(\frac{1-a^2y+\sqrt{(1-a^2)(1-a^2y^2)}}{2}\right)\biggl].
\end{split}
\end{equation}
To simplify this expression we define an effective ratio of modes which depends on $y$ as follows
\begin{equation}
    ay=\sqrt{4f_y(1-f_y)}\quad {\rm with}\quad f_y\in[0,1/2] \quad\Leftrightarrow\quad f_y=\frac{1-\sqrt{1-4f(1-f)y^2}}{2},
\end{equation}
which means for $f_1=f$ if $0\leq f\leq\frac{1}{2}$ and $f_0=0$.
For the latter we exploited the relation
\begin{equation}
    \sqrt{1-a^2}=\sqrt{1-4(1-f)f}=1-2f.
\end{equation}
A similar formula exists for $f_y$, namely $\sqrt{1-a^2y^2}=1-2f_y$. The quantities $f_y$ as well as $f$ can be understood as some measure quantifying the distance to the maximally mixed quantum state. This distance in terms of the von Neumann entropy is the biggest when $f=1/2$ and vanishes for $f=0$.

In this notation, the limiting von Neumann entropy of $\rho^{(A)}$ is
\begin{equation}\label{ent.max.result}
\begin{split}
    \lim_{N\to\infty}\frac{\braket{S_A}_y}{N}={}&f\log(2)-f-(1-f)\log\left[1-f\right]+(f-f_y)+\frac{1}{2}\log\left[\frac{1-f}{1-f_y}\right]\\
    &+\frac{1-2f}{2}\biggl[\frac{1+y}{2}\log\left(1-\biggl(\sqrt{f(1-f_y)}-\sqrt{f_y(1-f)}\biggl)^2\right)\\
    &+\frac{1-y}{2}\log\left(1-\biggl(\sqrt{f(1-f_y)}+\sqrt{f_y(1-f)}\biggl)^2\right)\biggl].
\end{split}
\end{equation}
The first term is the maximally possible entropy of system $A$ (the Page curve) while the second and third term are the influence that we consider fermionic Gaussian quantum states and thus a much smaller manifold compared to the original Hilbert space of all fermionic pure quantum states. The remaining terms are the impact that the initial quantum state $\rho$ of the whole system is not pure.
The limiting mutual information becomes
\begin{equation}\label{mut.max.result}
\begin{split}
    \lim_{N\to\infty}\frac{\braket{I_{AB}}_y}{N}={}&-2f_y-2(1-f)\log\left[1-f\right]+\log\left[\frac{1-f}{1-f_y}\right]\\
    &+\frac{1-2f}{2}\biggl[(1+y)\log\left(1-\biggl(\sqrt{f(1-f_y)}-\sqrt{f_y(1-f)}\biggl)^2\right)\\
    &+(1-y)\log\left(1-\biggl(\sqrt{f(1-f_y)}+\sqrt{f_y(1-f)}\biggl)^2\right)\biggl]\\
    &+f[(1+y)\log(1+y)+(1-y)\log(1-y)]
\end{split}
\end{equation}
because of~\eqref{mutual.max}. 

\begin{figure}[t]
    \centering
    \begin{tikzpicture}
    \draw (0,0) node[inner sep=0pt]{\includegraphics[width=\linewidth]{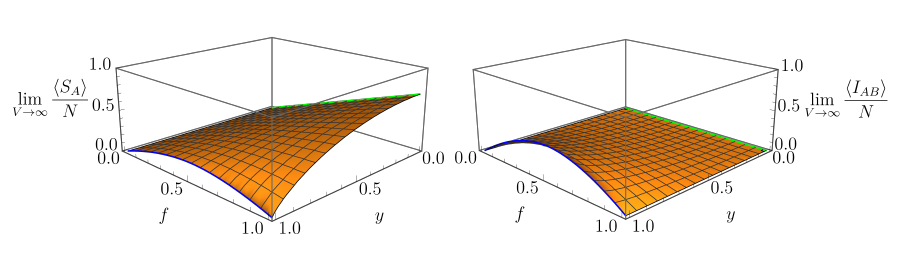}};
    \draw (-4.6,1.7) node[scale=.7]{\textbf{(a)}} (1.5,1.7) node[scale=.7]{\textbf{(b)}};
    \end{tikzpicture}
    \caption{Leading order of the average von Neumann entropy $\braket{S_A}=\braket{S(\rho^{(A)})}$ (plot \textbf{(a)}, see~\eqref{ent.max.result}) and mutual information $\braket{I_{AB}}=2\braket{S(\rho^{(A)})}-2N_As(y)$ (plot \textbf{(b)}, see~\eqref{mut.max.result}) for a large system in the maximally degenerate case. Both quantities are given as functions of the subsystem fraction $f=N_A/N$ and the singular value $y$ of the complex structure $J$ associated to the original quantum state $\rho$. We indicate the limiting functions for $y=1$ (blue) and $y=1$ (green).}
    \label{fig:vonneumann}
\end{figure}

The resulting mutual information for $f>1/2$ is given by replacing $f \leftrightarrow 1-f$ in~\eqref{mut.max.result}. For the von Neumann entropy~\eqref{ent.max.result} we make use of~\eqref{ent.rel.max} in such a case and express $\braket{S_A}$ in terms of $\braket{S_B}$ which now corresponds to the smaller system. We show the properly symmetrised results in Fig.~\ref{fig:vonneumann}.

The limit $y\to1$ to a pure quantum state, implying $f_y\to f$, reproduces the result in~\cite{bianchi2021page},
\begin{equation}
    \lim_{N\to\infty}\frac{\braket{S_A}_{y=1}}{N}=f(\log(2)-1)-(1-f)\log(1-f),
\end{equation}
while the mutual information becomes
\begin{equation}
     \lim_{N\to\infty}\frac{\braket{I_{AB}}_y}{N}=2\lim_{N\to\infty}\frac{\braket{S_A}_{y=1}}{N},
\end{equation}
\ie for a globally pure quantum state the mutual information is twice the entanglement entropy, which is well-known to be true.

The limit to the maximally mixed quantum state, realised by $y\to0$, results in  the leading order behaviour of the Page curve
\begin{equation}
    \lim_{N\to\infty}\frac{\braket{S_A}_{y=0}}{N}=f\log{2}\label{eq:sAy0}
\end{equation}
and the mutual information vanishes, $\lim_{N\to\infty}\braket{I_{AB}}_{y=0}/N=0$, as desired. This is again consistent with the well-known fact that a globally maximally mixed state does not have any correlations, neither classical nor quantum, so the mutual information must vanish.

\section{Summary and outlook}\label{sec:summary-outlook}

Using tools from random matrix theory and determinantal point processes, we extended previous studies of the average entanglement entropy of Haar-random pure Gaussian states to the average mutual information for mixed Gaussian states. For this goal, we described the level density of restricted complex structure $J_A$ in terms of the spectrum of the  complex structure $J$ for the whole system. 

Our methods also allow us to study the thermodynamic limit. As a proof of principle, we considered the simplest non-trivial case of mixed Gaussian states of maximal degeneracy, characterized by a complex structure $J$ with singular value $y\in[0,1]$. We derived closed formulas for the average von Neumann entropies of the reduced density operators in subsystems $A$ and $B$ as well as the average mutual information. In the limit $y\to 1$, we reproduce previously derived properties for pure states, namely the average mutual information being given by twice the average entanglement entropy, as expected. We are also able to capture the full transition from $y=0$ to $y=1$, as illustrated in Fig.~\ref{fig:vonneumann}, where the respective functional form is~\eqref{mut.max.result}. For the von Neumann entropy~\eqref{ent.max.result}, we could identify the respective contributions from the Page curve, the Gaussianity of the fermionic quantum states and the order of being mixed encoded in $y$ which allows an accessible interpretation.

In comparison, the limiting results for a general  complex structure $J$ of the whole system is highly involved, see~\eqref{entropy.limit} and~\eqref{mutual.limit}. We do not only derive the von Neumann entropy of the subsystems for the volume order but also for the constant one. These results are given in terms of a double integral which involve a saddle point solution. This solution can be represented implicitly as the solution of the algebraic equation~\eqref{saddle.2} or explicitly as the integral~\eqref{sol.2}. Since these results hold true for general singular values of the complex structure one can now start to study particular physical settings such as Gaussian fermionic quantum systems in an equilibrium. Such choices will yield particular limiting distributions for the singular values $y_j$ of $J$ that can be plugged into~\eqref{sol.2} and~\eqref{entropy.limit} to investigate how much generic complementary subsystems are correlated. Yet, the computation of the variance of the mutual information will pose a challenge, see~\cite{bianchi2021page,hackl-volume-law2022} for the variance and~\cite{2019Entrp..21..539W,2020JPhA...53g5302W,Huang:2021btj} higher cumulants for pure Gaussian fermionic quantum states. The reason is that we need to find the joint probability density of the singular values of the complex structures for both subsystems. Then, the random matrix results of~\cite{kieburg2019multiplicative} need to be extended.

Most of these results assume that the subsystem sizes $N_A$ and $N_B$ scale both linearly with the total number $N$ of fermionic modes. When, however, one of the two subsystems remains finitely sized in the limit $N\to\infty$, say, \eg, subsystem $A$, we have observed a stark simplification. As expected, the smaller of the two subsystems is close to being maximally mixed, see~\eqref{A.entropy.fixed.a}, while the larger of the two carries most of the entropy of the total system, though there is an order one correction to it as can be seen in~\eqref{B.entropy.fixed.e}. This leads to a mutual information of order one, {\it cf.}~\eqref{mut.finite.result}.

We would like to emphasise that our results even cover the case of the mutual information of an $L$-partite fermionic quantum system that is Gaussian. When only considering the mean value of the corresponding mutual information one can plug in the result~\eqref{entropy.limit} into
\begin{equation}
    \braket{I_{A_1,\ldots,A_L}}_y=\sum_{j=1}^L\braket{S(\rho^{(f_jN)}}_y-\sum_{k=1}^Ns(y_k),
\end{equation}
where $f_j\in[0,1]$ are the system fractions of the $L$ tensor components of the Hilbert space $\mathcal{H}$.

The results presented here can be seen as a direct generalization of the respective computations for pure states, which is typically phrased as choosing a bipartition into complementary subsystems $A$ and $B$ and then averaging over all pure Gaussian quantum states. However, an alternative perspective and interpretation of this setting is to fix a single pure quantum state in the system and then average over all possible system decompositions using the Haar measure on the group describing these decompositions. While Gaussian transformations technically form a double cover of the orthogonal group that governs the possible transformation of the Majorana modes, and thereby the possible decompositions by splitting into $(\gamma_1,\dots,\gamma_{2N_A})$ and $(\gamma_{2N_A+1},\dots,\gamma_{2N})$, this does not affect the average, as two Gaussian unitary transformations related to the same system decomposition will only differ by a complex phase without affecting the von Neumann entropy or the mutual information. Therefore, both, averaging over Gaussian unitary transformations or orthogonal system decompositions, require identical calculations and yield the same average. When moving to mixed states and mutual information, we can anew take both perspectives, either averaging over all (mixed) Gaussian states with fixed spectrum, as applying Gaussian unitary transformations will leave the spectrum invariant, or fixing a single mixed Gaussian state and then averaging over all orthogonal system decompositions.

There are several natural extensions of our work. First, one could consider to average over all Gaussian mixed states with an appropriate measure, thereby effectively replacing Gaussian unitary transformations by Gaussian channels, which can now also change the state's spectrum. Such a calculation would brake the previously described equivalence of averaging over states versus averaging over system decomposition present in our work. Second, one could consider general Haar distributed random unitary transformations including non-Gaussian ones, which would also preserve the state's spectrum, but map Gaussian to non-Gaussian (mixed) states. Third, many of our considerations could be repeated for other measures of correlations, such a negativity~\cite{plenio2005logarithmic} or entanglement of formation~\cite{horodecki2009quantum}. While these quantities are generally more challenging to compute compared to the expected mutual information, their potential advantage is that they are able to clearer distinguish quantum from classical correlations. Forth, one could also consider non-fermionic systems and in particular bosonic system, in which case different entanglement measures (such as negativity) may behave vastly differently to the case of fermions~\cite{shapourian2017partial}.

\section*{Acknowledgment}
MK would like to devote this work to Santosh Kumar who also worked on random matrix ensembles applied to quantum information problems and who recently passed away at a very young age.

MK is partially funded by the Australian Research Council via
the Discovery Project grant DP250102552, by the Alexander-von-Humboldt Foundation and by the Swedish Research Council under the grant no. 2021-06594 while the author was in residence at the Mittag-Leffler Institute in Djursholm, Sweden during November/December of 2024. He is also grateful to the Faculty of Physics of the University Duisburg-Essen  for their hospitality during his sabbatical.

LH acknowledges support by the Alexander von Humboldt Foundation, by grant $\#$62312 from the John Templeton Foundation, as part of the \href{https://www.templeton.org/grant/the-quantuminformation-structure-ofspacetime-qiss-second-phase}{‘The Quantum Information Structure of Spacetime’ Project (QISS)}, by grant $\#$63132 from the John Templeton Foundation and an Australian Research Council Australian Discovery Early Career Researcher Award (DECRA) DE230100829 funded by the Australian Government. The opinions expressed in this publication are those of the authors and do not necessarily reflect the views of the respective funding organization.

JM would like to dedicate this work to his parents Elena Aguas and Rodrigo Maldonado, for all their unconditional support during his studies. He also thanks his supervisors Dr. Mario Kieburg and Dr. Lucas Hackl, for their support during his Master's studies ensuring his academic and personal growth. The present work is based on this Master's project.

\section*{References}
\bibliography{references.bib}
\bibliographystyle{iopart-num}

\end{document}